\documentclass{book}
\usepackage{setspace}
\usepackage[utf8]{inputenc}
\usepackage{authblk}
\usepackage[hidelinks]{hyperref}
\usepackage{graphicx}
\usepackage{amsmath}
\usepackage{algorithmic}
\usepackage{algorithm}
\usepackage{amsmath}
\usepackage{mathptmx}
\usepackage{enumitem}
\usepackage{framed}

\begin{document}

	\title{Computer simulations in science and engineering\\Concepts - Practices - Perspectives}
	
	
	\author{Juan M. Dur\'an\footnote{Delft University of Technology}\\\textit{Springer - The Frontiers Collection}\\ ISBN: 978-3-319-90880-9\\\vspace{10mm}\textbf{Download the rest of the book: juanmduran.net}}

\date{}
\maketitle

\tableofcontents

\noindent To mom, dad, and Jo\\
\noindent For their unconditional support and love.\\

\vspace{1cm}

\noindent To my Bee\\
\noindent I could have not made this journey without you.

\vspace{1cm}

\noindent To Mauri\\
\noindent In true friendship.

\vspace{1cm}

\noindent To Manuel\\
\noindent With love and admiration.

\section*{Preface}

The ubiquitous presence of computer simulations in all kinds of research areas evidence their role as the new driving force for the advancement of science and engineering research. Nothing seems to escape the image of success that computer simulations project onto the research community and the general public. One simple way to illustrate this consists in asking ourselves how would contemporary science and engineering look like without the use of computer simulations. The answer would certainly diverge from the current image we have of scientific and engineering research.

As much as computer simulations are successful, they are also methods that fail in their purpose of inquiring about the world; and as much as researchers make use of them, computer simulations raise important questions that are at the heart of contemporary science and engineering practice. In this respect, computer simulations make a fantastic subject of research for the natural sciences, the social sciences, engineering and, as in our case, also for philosophy. Studies on computer simulations touch upon many different facets of scientific and engineering research, and evoke philosophically inclined questions of interpretation with close ties to problems in experimental settings and engineering applications.
 
  
This book will introduce the reader, in an accessible and self-contained manner, to these various fascinating aspects of computer simulations. An historical study on the conceptualization of computer simulations throughout the past sixty years opens up the vast world of computer simulations and their implications. The focus then is shifted to the discussion on their methodology, their epistemology, and the  possibilities of an ethical framework, among other issues.
 
 The scope of this book is relatively broad in order to familiarize the reader with the many facets of computer simulations. Throughout the book, I have sought to maintain a healthy balance between the conceptual ideas associated with the philosophy of computer simulations on the one hand, and their practice in science and engineering on the other hand. To this end, the book has been conceived for a broad audience, from scientists and engineers, policy makers and academics, to the general public. It welcomes anyone interested in philosophical questions -- and conceivable answers -- to issues raised by the theory and practice of computer simulations. It must be mentioned that, although the book is written in a philosophical tone, it does not engage in deep philosophical discussions. Rather, it seeks to explore the synergy between technical aspects of computer simulations and the philosophical value there emerging. In this respect, the ideal readers of this book are researchers across disciplines working on computer simulations but holding philosophical inclinations. This is, of course, not to say that professional philosophers would not find in its pages problems and questions for their own research.
 
 One beautiful thing about computer simulations is that they offer a fertile field of research, both for researchers using the simulations as well as those reflecting upon them. In this respect, although the book might have some merits, it also falls short in many respects. For instance, it does not address the work of computer simulations in the social sciences, a very fruitful area of research. It also does not discuss the use of computer simulations in and for policy making, their uses for reporting to the general public, nor their role in a democratic society where science and engineering practice is a common good. This is certainly unfortunate. But there are two reasons that, I hope, excuse the book from these shortcomings. One is that I am not a specialist in any of these fields of research, therefore my contribution would have been of little interest. Each of the fields mentioned brings about specific issues in their own right that those involved in their study know best. The second reason stems from the fact that, as all researchers know, time -- and also in this case, space -- are tyrant. It would be an impossible task to even scratch the surface of the many areas where computer simulations are active and thriving. 
 
As a general rule for the book, I present a given topic and discuss problems and potential solutions to it. No topic should be addressed as unrelated to any other topic in the book, nor should a proposed answers be taken as final. In this sense, the book aims at motivating further discussions, rather than providing a closed set of topics and the answers to their core issues. Each chapter should nevertheless present a self-contained discussion of a general theme of computer simulations. I must also mention that each chapter contains profuse references to the specialized literature, giving the reader the opportunity to pursue further his or her own interests on a given subject.


The book is organized as follows. In chapter 1, I address the question `what are computer simulations?' by giving an historical overview of the concept. Tracking back the concept of computer simulation to the early 1960s, we will soon realize that many contemporary definitions owe much to these early attempts. A proper grasp of the history of the concept will turn out to be very important for the development of a solid understanding of computer simulations. In particular, I identify two traditions, one that puts the emphasis on implementing mathematical models on the computer, and another for which the yellow prominent feature is the representational capacity of the computer simulation. Depending on which tradition researchers chose to follow, the assumptions and implications to be drawn from computer simulations will differ. The chapter ends with a discussion on the now standard classification of computer simulations.

The core of chapter 2 is to introduce and discuss in detail the constituents of \textit{simulation models} -- that is, the models at the basis of computer simulations. To this end, I discuss diverse approaches to scientific and engineering models with the purpose of entrenching simulation models as a rather different kind. Once this is accomplished, the chapter goes on presenting and discussing three units of analysis constitutive of computer simulations, namely, the specification, the algorithm, and the computer process. This chapter is the most technical of the book, as it draws extensively from studies on software engineering and computer science. In order to balance this with some philosophy, it also presents several problems related to these units of analysis -- both individually and in relation to each other.

The sole purpose of chapter 3 is to present the discussion on whether computer simulations are epistemologically equivalent to laboratory experimentation. The importance of establishing such equivalence has its roots in a tradition that takes experimentation as the solid foundation for our insight into the world. Since much of the work demanded from computer simulations is to provide knowledge and understanding of real-world phenomena that would otherwise not be possible, then the question of their epistemological power in comparison with laboratory experimentation naturally occurs. Following the philosophical tradition of discussing these issues, I focus on the now time-honored problem of the `materiality' of computer simulations.

Although chapters 4 and 5 are independent of each other, they do share the interest in establishing the epistemological power of computer simulations. While chapter 4 does so by discussing the many ways in which computer simulations are reliable, chapter 5 does it by showing the many epistemic functions attached to computer simulations. These two chapters, then, represent my contribution to the many attempts to ground the epistemic power of computer simulations. Let us note that these chapters are, at their basis, an answer to chapter 2, which discusses computer simulations \textit{vis \`a vis} laboratory experimentation.
 
Next, chapter 6 addresses issues that are arguably less visible in the literature on computer simulations. The core question here is whether computer simulations should be understood as a third paradigm of scientific and engineering research -- theory, experimentation, and Big Data being the first, second, and fourth paradigm respectively. To this end, I first discuss the use of Big Data in scientific and engineering practice, and what it means to be a paradigm. With these elements in mind, I begin a discussion on the possibilities of holding causal relations in Big Data science as well as computer simulations, and what this means for the establishment of these methodologies as paradigms of research. I finish the chapter with a comparison between computer simulations and Big Data with a special emphasis on what sets them apart.

The last chapter of the book, chapter 7, addresses an issue that has been virtually unexplored in the literature on ethics of technology, that is, the prospect of an ethics of computer simulations. Admittedly, the literature on computer simulations is more interested in their methodology and epistemology, and much less on the ethical implications that comes with designing, implementing, and using computer simulations. In response to this lack of attention, I approach this chapter as an overview to the ethical problems addressed in the specialized literature.

\vspace{\baselineskip}
\begin{flushright}\noindent
	Stuttgart, Germany,\hfill {Juan M. Dur\'an}\\
	
\end{flushright}

\section*{Acknowledgements}

As is usually the case, there are many people who have contributed in making this book possible. First and foremost, I would like to thank Marisa Velasco, P\'io Garc\'ia and Paul Humphreys for their initial encouragement to write this book. All three have had a strong presence in my education as a philosopher, and this book certainly owns them much. To all three, my gratitude.

This book started in Argentina and ended in Germany. As a postdoc at the Centro de Investigaciones de la Facultad de Filosof\'ia y Humanidades (CIFFyH), Universidad Nacional de C\'ordoba (UNC - Argentina), funded by the National Scientific and Technical Research Council (CONICET), I had the chance to write and discuss the first chapters with my research group. For this, I am grateful to V\'ictor Rodr\'iguez, Jos\'e Ahumada, Juli\'an Reynoso, Maximiliano Bozzoli, Penelope Lodeyro, Xavier Huvelle, Javier Blanco, and Mar\'ia Silvia Polzella. Andr\'es Il\v{c}i\'c is another member of this group, but he deserves special recognition. Andr\'es read every chapter of the book, made thoughtful comments, and amended several mistakes that had gone unnoticed by me. He has also diligently checked many of the formulas I use in the book. For this and for the innumerable discussions we have had, thank you Andy. Naturally, all mistakes are my responsibility. A very sincere thanks goes to CIFFyH, the UNC, and CONICET, for their support to the Humanities in general and to me in particular. 

I also have many people to thank that, although they did not contribute directly to the book, they showed their support and encouragement throughout good and bad days. My eternal gratitude goes to my good friend Mauricio Zalazar, and my sister Jo. Thank you guys for being there when I need you the most. My parents have also been a constant source of support and love, thank you mom and dad, this book would not exist without you. Thanks also go to Víctor Scarafía and my newly adopted family: the Pompers \& the Ebers. Thank you all guys for being so great. Special thanks go the the two Omas: Oma Pomper and Oma Eber. I love you ladies. Finally, thanks go to Peter Ostritsch for his support in many aspect of my life, most of them unrelated to this book.

The book ended at the Department of Philosophy of Science and Technology of Computer Simulation, at the High-Performance Computing Center Stuttgart (HLRS), University of Stuttgart. The department was created by Michael Resch and Andreas Kaminski and funded by the Ministerium f\"ur Wissenschaft, Forschung und Kunst Baden-W\"urttemberg (MWK), whom I thank for providing a comfortable atmosphere for working on the book. I extend my thanks to all the members of the department, Nico Formanek, Michael Hermann, Alena Wackerbarth, and Hildrun Lampe, I will never forget the many fundamental philosophical discussions we had at lunch -- and over our new espresso machine -- on the most varied of topics. I feel especially lucky to share an office with Nico and Michael, good friends and great philosophers. Thank you guys for checking the formulas that I included in the book. The mistakes are, again, entirely my responsibility. Thanks also go to Bj\"orn Schembera, a true philosopher in disguise, for our time discussing so many technical issues about computer simulations, some of which found a place in the book. From the visualization department at the HLRS, thanks go to Martin Aum\"uller, Thomas Obst, Wolfgang Schotte, and Uwe Woessner who patiently explained to me the many details of their work as well as provided the images for Augmented Reality and Virtual Reality discussed in the chapter on visualization. The images of the tornado that are also in that chapter were provided by the National Center for Supercomputing Applications, University of Illinois at Urbana-Champaign. For this, I am in great debt to Barbara Jewett for her patience, time, and great help in finding the images.

 I would also like to express my gratitude to my former PhD advisor, Ulla Pompe-Alama for encouragement and suggestions on early drafts. Special thanks go to Raphael van Riel and the University of Duisburg-Essen for their support on a short term fellowship. For several reasons, directly and indirectly related to the book, I am in debt with Mauricio Villase\~nor, Jordi Valverd\'u, Leandro Giri, Ver\'onica Pedersen, Manuel Barrantes, Itat\'i Branca, Ram\'on Alvarado, Johannes Lenhard and Claus Beisbart. Thank you all for your comments, suggestions, encouragement during the different stages of the book, and for having me all sorts of philosophical conversations with me. Angela Lahee, my editor at Springer, deserves much credit for her patience, encouragement, and helpful support in the production of this book. Although I would have kept on polishing the ideas in this book and, equally important, my English, the time to put an end to it is upon me.
 
My enormous gratitude goes to Tuncer \"Oren, with whom I shared many correspondences on ethical and moral problems of computer simulations, leading to the final chapter of the book. Prof. \"Oren's love and dedication to philosophical studies on computer simulation is a source of inspiration. 

Finally, in times where science and technology are undoubtedly a fundamental tool for the progress of society, it is heartbreaking to see how the current government of Argentina -- and in many other places in Latin-America as well -- are cutting funding in science and technology research, humanities, and the social sciences. I observe with equal horror the political decisions explicitly targeted to the destruction of the educational system. I then dedicate this book to the Argentinean scientific and technological community, for they have shown time and time again their greatness and brilliance despite unfavorable conditions.\\

This book owes Kassandra too much. She left her impression when correcting my English, when suggesting me to re-write a whole paragraph, and when she let go a planned appointment that I forgot while finishing a section. For this, and for thousands of other reasons, this book is entirely dedicated to her.

\section*{Introduction}\label{Introduction}
\label{chap:Introduction} 
\addcontentsline{toc}{chapter}{Introduction}

In 2009, a debate erupted around the question of whether computer simulations introduce novel \textit{philosophical} problems or if they are merely a \textit{scientific} novelty. Roman Frigg and Julian Reiss, two prominent philosophers that ignited the debate, noted that philosophers have largely assumed some form of philosophical novelty of computer simulations without actually engaging the question of its possibility. Such an assumption rested on one simple confusion: philosophers were thinking that scientific novelty licenses philosophical novelty. This gave course to issuing a warning over the growth of overemphasized and generally unwarranted claims about the philosophical importance of computer simulations. This growth, according to the authors, was reflected in the increasing number of philosophers convinced that the philosophy of science, nourished by computer simulations, required an entirely new epistemology, a revised ontology, and novel semantics. 

It is important to point out that Frigg and Reiss are not objecting to the novelty of computer simulations in scientific and engineering practice, nor their importance in the advancement of science, but rather that simulations raise few, if any, new philosophical question. In their own words, ``[t]he philosophical problems that do come up in connection with simulations are not specific to simulations and most of them are variants of problems that have been discussed in other contexts before. This is not to say that simulations do not raise new problems of their own. These specific problems are, however, mostly of a mathematical or psychological, not philosophical nature'' \cite[595]{Frigg2009a}. 

I share Frigg and Reiss' puzzlement on this issue. It is hard to believe that a \textit{new} scientific method -- instrument, mechanism, etc. -- however powerful as it might be, could all by itself imperil current philosophy of science and technology to the point that they need to be rewritten. But this is only true if we accept the claim that computer simulations come to \textit{rewrite} long-standing disciplines, which I do not think it is the case. To me, if we are able to reconstruct and give new meaning to old philosophical problems in light of computer simulations, then we are basically establishing their philosophical \textit{novelty}. 

Let us now ask the question in what sense are computer simulations a philosophical novelty? There are two ways to unpack the problem. Either computer simulations pose a series of philosophical questions that escape standard philosophical treatment, in which case they can be added to our philosophical corpus; or they challenge established philosophical ideas, in which case the current corpus expands standard debates into new domains. The first case has been proposed by \cite{Humphreys2009}, whereas the second case has been argued by myself \cite{Duranunderreview}. Let me now briefly discuss why computer simulations represent, in many respects, a scientific and philosophical novelty.

The core of Humphreys' argument is to recognize that we could either understand computer simulations by focusing on how traditional philosophy illuminates their study (e.g., through a philosophy of models, or a philosophy of experiment), or by focusing exclusively on aspects about computer simulations that constitute, in and by themselves, genuine philosophical challenges. It is this second way of looking at the questions about their novelty that grants philosophical importance to computer simulations. 

The chief claim here is that computer simulations can solve otherwise intractable models and thus amplify our cognitive abilities. But such amplification comes with a price ``for an increasing number of fields in science, an exclusively anthropocentric epistemology is no longer appropriate because there now exist superior, non-human, epistemic authorities'' \cite[617]{Humphreys2009}. Humphreys calls this the \emph{anthropocentric predicament} as a way to illustrate current trends in science and engineering where computer simulations are moving humans away from the center of production of knowledge. According to him, a brief overview on the history of philosophy of science shows that humans have always been at the center of production of knowledge. This conclusion includes the period of the logical and empirical positivism, where the human senses were the ultimate authority \cite[616]{Humphreys2009}. A similar conclusion follows from the analysis of alternatives to the empiricist, such as Quine's and Kuhn's epistemologies. 

When confronted with claims about the philosophical novelty of computer simulations, Humphreys points out that the standard empiricist viewpoint has prevented a complete separation between humans and their capacity to evaluate and produce scientific knowledge. The anthropocentric predicament, then, comes to highlight precisely this separation: it is the claim that humans have lost their privileged position as the ultimate epistemic authority.\footnote{Humphreys makes a further distinction between scientific practice completely carried out by computers -- one that he calls \textit{the automated scenario} -- and one in which computers only partially fulfill scientific activity -- that is, the \textit{hybrid scenario}. He restricts his analysis, however, to the hybrid scenario \cite[616-617]{Humphreys2009}.} The claim finally gets its support from the view that scientific practice only progresses because new methods are available for handling large amounts of information. Handling information, according to Humphreys, is the key for the progress of science today, which can only be attainable if humans are removed from the center of the epistemic activity \cite[8]{Humphreys2004}.

The anthropocentric predicament, as philosophically relevant as it is in itself, also brings about four extra novelties unanalyzed by the traditional philosophy of science. Those are \textit{epistemic opacity}, \textit{the temporal dynamics of simulations}, \textit{semantics}, and the \textit{in practice/in principle} distinction. All four are novel philosophical issues brought up by computer simulations; all four have no answer in traditional philosophical accounts of models and experimentation; and all four represent a challenge for the philosophy of science. 

The first novelty is \textit{epistemic opacity}, a topic that is currently attracting much attention from philosophers. Although I discuss this issue in some detail in section \ref{Epistemic_opacity}, briefly mentioning the basic assumptions behind epistemic opacity will shed some light on the novelty of computer simulations. Epistemic opacity, then, is the philosophical position that takes that it is impossible for any human to know all the epistemically relevant elements of a computer simulation. Humphreys presents this point in the following way: ``A process is essentially epistemically opaque to [a cognitive agent] X if and only if it is impossible, given the nature of X, for X to know all of the epistemically relevant elements of the process'' \cite[618]{Humphreys2009}. To put the same idea in a different form, if a cognitive agent could stop the computer simulation and take a look inside, she would not be able to know the previous states of the process, reconstruct the simulation up to the point of stop, or predict future states given previous states. Being epistemically opaque means that, due to the complexity and speed of the computational process, no cognitive agent could know what makes a simulation an epistemically relevant process.

A second novelty that is related to epistemic opacity is the \textit{temporal dynamics} of computer simulations. This concept has two possible interpretations. Either it refers to the necessary computer-time to solve the simulation model, or it stands for the temporal development of the target system as represented in the simulation model. A good example that merges these two ideas is a simulation of the atmosphere: the simulation model represents the dynamics of the atmosphere for a year and it takes, say, ten days to compute.

These two novelties nicely illustrate what is typical of computer simulations, namely, the inherent complexity of simulations in themselves, as is the case of epistemic opacity and the first interpretation of temporal dynamics; and the inherent complexity of the target systems that computer simulations usually represent, as it is the case of the second interpretation of temporal dynamics. What is common between these two novelties is that they both entrench computers as the epistemic authority since they are able to produce reliable results that no human or group of human could produce by themselves. Either because the process of computing is too complex to follow or because the target system is too complex to comprehend, computers become the exclusive source for obtaining information about the world. 

The second interpretation of temporal dynamics is tailored to the novelty of the \textit{semantics}, which asks the question of how theories and models represent the world, now adjusting the picture to fit a computer algorithm. Thus, the chief issue here is how the syntax of a computer algorithm maps onto the world, and how a given theory is actually brought into contact with data. 

Finally, the distinction in principle/in practice is intended to sort out what is applicable in practice and what is applicable only in principle. To Humphreys, it is a philosophical fantasy to say that, in principle, all mathematical models find a solution within computer simulations \cite[623]{Humphreys2009}. It is a fantasy because it is clearly false, although philosophers have claimed its possibility -- hence, in principle. Humphreys suggests, instead, that in approaching computers, philosophers must keep a more down-to-earth attitude, limited to the technical and empirical constraints that simulations can offer.

My position is complementary to Humphreys' in the sense that it shows how computer simulations challenge established ideas in the philosophy of science. To this end, I begin by arguing for a specific way of understanding simulation models, the kind of model at the basis of computer simulations. To me, a simulation model \textit{recasts} a multiplicity of models into one `super-model'. That is to say, simulation models are an amalgam of different sorts of  computer models, all having their own scales, input parameters, and protocols. In this context, I claim for three novelties in philosophy, namely, \textit{representation}, \textit{abstraction}, and \textit{explanation}. 

About the first novelty, I claim that the multiplicity of models implies that \textit{representation} of a target system is more holistic in the sense that it encompasses all and every model implemented in the simulation model. To put the same idea in a rather different form, the representation of the simulation model is not given by any individual implemented model but rather by the combination of all of them. 

The challenge that computer simulations bring to the notion of \textit{abstraction} and idealization is that, typically, the latter presupposes some form of \textit{neglecting} stance. Thus, \textit{abstraction} aims at ignoring concrete features that the target system possesses in order to focus on their formal set-up; \textit{idealizations}, on the other hand, come in two flavors: while Aristotelian idealizations consist in `stripping away' properties that we believe not to be relevant for our purposes, Galilean idealizations involve deliberate distortions. Now, in order to implement the required variety of models into a single simulation model, it is important to count on techniques by which information is hidden from the users, but not neglected from the models \cite{Colburn2007}. This is to say that the properties, structures, operations, relations and the like present in each mathematical model can be effectively implemented into the simulation model without stating explicitly how such implementation is carried out. 

Finally, \textit{scientific explanation} is a time-honored philosophical topic where much has been said. When it comes to explanation in computer simulations, however, I propose a rather different look at the issue than the standard treatment offers. One interesting point here is that, in the classic idea that explanation is of a real-world phenomenon I oppose the claim that explanation is, first and foremost, of the results of computer simulations. In this context, many new questions emerge seeking for an answer. I discuss scientific explanation in more detail in section \ref{Explanatory_force}.\\

As I have mentioned before, I do believe that computer simulations raise novel questions for the philosophy of science. This book is living proof of that belief. But even if we do not believe in their philosophical novelty, we still need to understand computer simulations as scientific novelties with a critical and philosophical eye. To these ends, this book presents and discusses several theoretical and philosophical issues at the heart of computer simulations. Having said all of this, we may now submerge ourselves into their pages.

\chapter{The universe of computer simulations}\label{chap:The_universe_of_CS}
	
The universe of computer simulations is vast, flourishing in almost every scientific discipline, and still resisting a general conceptualization. From the early  computations of the Moon's orbit carried out by punched card machines, to the most recent attempts to simulate quantum states, computer simulations have a uniquely short but very rich history.

We can situate the first use of a machine for scientific purposes in England at the end of the 1920s. More precisely, it was in 1928 when the young astronomer and pioneer in the use of machines Leslie J. Comrie predicted the motion of the Moon for the years 1935 to 2000. During that year, Comrie made intensive use of a Herman Hollerith punched card machine to compute the summation of harmonic terms in predicting the Moon's orbit. Such groundbreaking work would not stay in the shadows, and by the mid 1930s it had cross the ocean to Columbia University in New York City. It was there that Wallace Eckert founded a laboratory that made use of punched card tabulating machines -- now built by IBM -- to perform calculations related to astronomical research, including of course an extensive study of the motion of the Moon. 

Both Comrie's and Eckert's uses of punched card machines share a few commonalities with today's use of simulations. Most prominently, both implement a special kind of model that describes the behavior of a target system, and which can be interpreted and computed by a machine. While Comrie's computing rendered data about the motions of the Moon, Eckert's simulation described planetary movement. 

These methods certainly pioneered and revolutionized their respective fields, as well as many other branches of the natural and social sciences. However, Comrie's and Eckert's simulations significantly differ from today's \textit{computer simulations}. Upon closer inspection, differences can be found everywhere. The introduction of silicon based circuits, as well as the subsequent standardization of the circuit board, made a significant contribution to the growth of computational power. The increase in the speed of calculation, size of memory, and expressive power of programming language forcefully challenged the established ideas on the nature of computation and of its domain of application. Punched card machines rapidly became obsolete as they are slow in speed, unreliable in their results, limited in their programming, and based on stiff technology (e.g., there were very few exchangeable modules). In fact, a major disadvantage of the punched card over modern computers is that they are error-prone and time-consuming machines, and therefore the reliability of their results as well as their representational accuracy is difficult to ground. However, perhaps the most radical difference between Comrie's and Eckert's simulations, on the one hand, and modern computer simulations on the other, is the automation process that characterizes the latter. In today computer simulations, researchers are losing grounds on their influence and power to interfere in the process of computing, and this will become more prominent as complexity and computational power increases. 

Modern computers come to amend many aspect of scientific and engineering practice with more precise computations, and more accurate representations. Accuracy, computational power, and reduction of errors are, as we will see, the main keys of computer simulations that unlock the world. 

In light of contemporary computers, then, it is not correct to maintain that Comrie's prediction of the motion of the Moon and Eckert's solution of planetary equations are computer simulations. This is, of course, not to say that they are not simulations at all. But in order to accommodate to the way scientists and engineers use the term today, it is not sufficient to be able to compute a special model or to produce certain kinds of results about a target system. Speed, storage, language expressiveness, and the capacity to be (re)programmed are chief concepts for the modern notion of computer simulation. 

What are computer simulations then? This is a philosophically motivated question that has found different answers from scientists, engineers, and philosophers. The heterogeneity of their answers makes explicit how differently each researcher conceives computer simulations, how their definitions vary from one generation to the next, and how difficult it is to come up with a unified notion. It is important, however, to have a good sense of their nature. Let us discuss this in more extent.

\section{What are computer simulations?}\label{sec:What-are-CS?}

Recent philosophical literature takes computer simulations as aids for overcoming imperfections and limitations of human cognition. Such imperfections and limitations are tailored to the natural human constrains of computing, processing and classifying large amounts of data. Paul Humphreys, one of the first contemporary philosophers to address computer simulations from a purely philosophical viewpoint, takes then as an `amplification instrument,' that is, one that speeds up what the unaided human could not do by herself \cite[110]{Humphreys2004}. In a similar sense, Margaret Morrison, yet another central figure in philosophical studies on computer simulations, considers that although they are another form of modeling, ``given the various functions of simulation [...] one could certainly characterize it as a type of `enhanced' modelling'' \cite[47]{Morrison2009}. 

Both claims are fundamentally correct. Computer simulations compute, analyze, render, and visualize data in many ways that are unattainable for any group of humans. Contrast, for instance, the time required for a human to identify potential antibiotics for infections diseases such as anthrax, with a simulation of the ribosome in motion at atomic detail \cite{Laboratory2015}. Or, if preferred, compare any set of human computational capabilities with the supercomputers used at the High Performance Computing Center Stuttgart, home of the Cray XC40 Hazel Hen with a peak performance of 7.42 Petaflops and a memory capacity of 128 GB per node.\footnote{It is worth noting that our neuronal network activity is, in some specific cases, faster than any supercomputer. According to relatively recent publication, Japan's \emph{Fujitsu K} computer, consisting of 82,944 processors, takes about 40 minutes to simulate one second of neuronal network activity in real, biological time. In order to partially simulate the human neural activity, researchers create about 1.73 billion virtual nerve cells that were connected to 10.4 trillion virtual synapses \cite{Himeno2013}.} 

As pointed out by Humphreys and Morrison, there are different senses in which computer simulations enhance our capacities. This could be by amplifying our calculation skills, as Humphreys suggests, or it could be by enhancing our modeling abilities, as Morrison suggests.

One would be naturally inclined to think that computer simulations amplify our computational capability as well as enhance our modeling abilities. However, a quick look at the history of the concept shows otherwise. To some authors, a proper definition must highlight the importance of finding solutions to a model. To others, the right definition centers the attention to describing patterns of behavior of a target system. Under the first interpretation, the computational power of the machine allows us to solve models that, otherwise, would be analytically intractable. In that respect, a computer simulation `amplifies' or `enhances' our cognitive capacities by providing computational power to what is beyond our cognitive reach. The notion of computer simulation is then dependent on the physics of the computer and furnishes the idea that technological change expands the boundaries of scientific and engineering research. Such a claim is also historically grounded. From Hollerith's punched card machines to the silicon-based computer, the increment of the physical power of computers has enabled scientists and engineers to find different solutions to a variety of models. Let me call this first interpretation \textit{the problem-solving viewpoint} on computer simulations. 

Under the second interpretation, the emphasis is on the capacity of the simulation to describe a target system. For this, we have a powerful language that represents, to certain acceptable degrees of detail, several levels of description. In that respect, a computer simulation `amplifies' or `enhances' our modeling abilities by providing more accurate representation of a target system. Thus understood, the notion of computer simulation is tailored to the way in which they describe a target system, and thus on the computer language used, modularization methods, software engineering techniques, etc. I call this second interpretation the \textit{description of patterns of behavior} viewpoint on computer simulations. 

Because both viewpoints emphasize different -- although not necessarily incompatible -- interpretations of computer simulations as enhancers, some distinctions can be drawn. For starters, under the problem-solving viewpoint, computer simulations are not experiments in any traditional sense, but rather the manipulation of an abstract and formal structure (i.e., mathematical models). In fact, to many advocates of this viewpoint, experimental practice is confined to the traditional laboratory as computer simulations are more of a crunching numbers practice, closer to mathematics and logic. The description of patterns of behavior viewpoint, on the other hand, allows us to treat computer simulations as experiments in a straightforward sense. The underlying intuition is that by means of describing the behavior of a target system, researchers are capable of carrying out something very similar to traditional experimental practice, such as measuring values, observing quantities, and detecting entities. 

Understanding things this way has some kinship with the methodology of computer simulations. As I discuss later, the problem-solving technique viewpoint considers a simulation to be the \textit{direct} implementation of a model on a physical computer. That is to say, mathematical models are implemented on the computer \textit{simpliciter}. The description of patterns of behavior viewpoint, instead, holds that computer simulations have a proper methodology which it is rather different from anything we have seen in the scientific and engineering arena. These methodological differences between the two viewpoints turn out to be central for later disputes about the novelty of computer simulations in scientific and engineering research. 

Another difference between these two viewpoints lies in the reasons for using computer simulations. Whereas the problem-solving viewpoint asserts that the use of computer simulations is only pragmatically justified when the model cannot be solved by more traditional methods, the description of patterns of behavior viewpoint considers that computer simulations offer valuable insight into the target system despite its analytical intractability. Let us note that what is also at stake here is the epistemic priority of one method over another. If the use of computer simulations is only justified when the model cannot be analytically solved -- as many advocates of the problem-solving viewpoint claim -- then analytic methods are epistemically superior to computational. This configures a specific standpoint regarding the place that computer simulations have in the scientific and engineering agenda. In particular, it significantly downplays the reliability of computer simulations for research in uncharted territory. We will have more to say about this throughout this book.\footnote{Many philosophers have tried to understand the nature of computer simulations. What I have offered above is just one possible characterization. For more, the reader could refer to the following authors \cite{Winsberg2010, Vallverdu2014, Morrison2015, Winsberg2015, Saam2016}.}

Finally, tackling problems in computer simulations can be very different depending on the viewpoint adopted. For the problem-solving viewpoint, any issues related to the results of the simulation (e.g., accuracy, computability, representability, etc.) can be solved on technical grounds (i.e., by increasing speed and memory, changing the underlying architecture, etc.). Instead, for the advocate of the description of patterns of behavior, the same issues have an entirely different treatment. Incorrect results, for instance, are approached by analyzing practical considerations at the design level, such as new specifications for the target system, alternative assessments of expertise knowledge, new programming languages, etc. In the same vein, incorrect results might be due to misrepresentation of the target system at the design, specification, and programming stages (see section \ref{Computer_simulations}), or misrepresentations at the computational stage (e.g., errors during computing time  -- see section \ref{Error_and_uncertainty}). The researcher's general understanding as well as solution to these issues changes significantly depending on the viewpoint adopted.

\label{example_Woolfson_Pert} 
To illustrate some of the points made so far, take a simple equation-based computer simulation of the dynamics of satellite orbiting around a planet under tidal stress. To simulate such dynamics, researchers typically begins with a mathematical model of the target system. A good model is furnished by classical Newtonian mechanics, as described by M. M. Woolfson and G. J. Pert in \cite{Woolfson1999}. 

For a planet of mass $M$ and a satellite of mass $m$ $(\ll M)$, in an orbit of semi-major axis $a$ and eccentricity $e$, the total energy is

\begin{equation} 
E=-\frac{GMm}{2a} \label{eq:Formula 1.31a}
\end{equation}

and the angular momentum is

\begin{equation} 
H=\{GMa(1-e^{2})\}m  \label{eq:Formula 1.31b}
\end{equation}

\begin{figure}
	\begin{center}
		\includegraphics[scale=0.4]{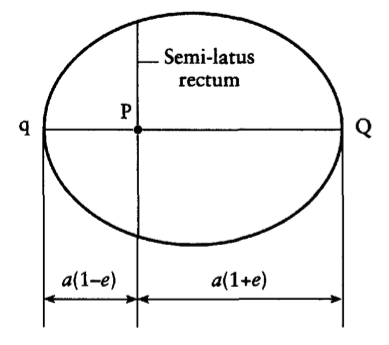}
		\caption{The elliptical orbit of a satellite relative to the planet at one focus. Points $q$ and $Q$ are the nearest and furthest points from the planet, respectively. \cite[19]{Woolfson1999}}
		\label{fig:The_elliptical_orbit_of_a_satellite}		
\end{center}
\end{figure}

\begin{figure}
	\begin{center}
		\includegraphics[scale=0.4]{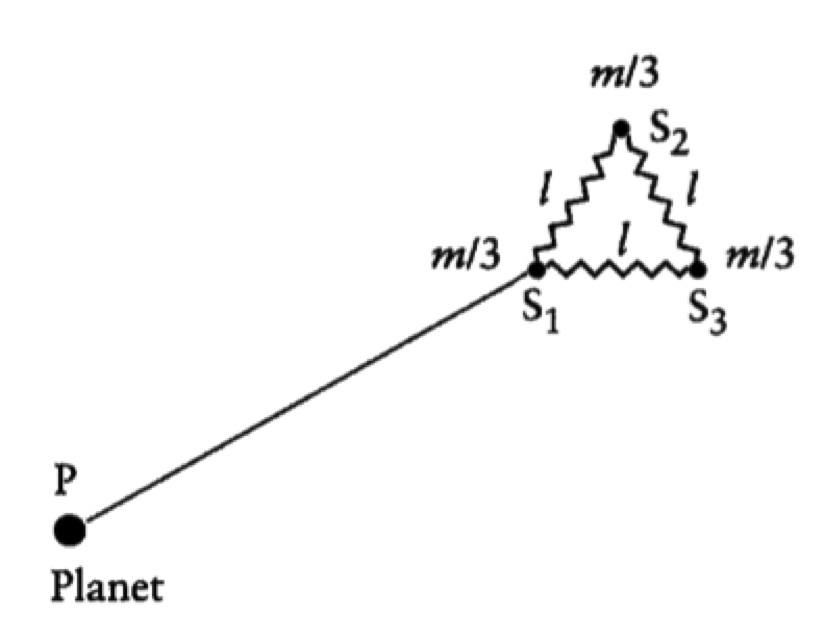}
		\caption{The satellite is described by three masses, each $m/3$, connected by springs each of the same unstrained length, $l$. \cite[19]{Woolfson1999}}
		\label{fig:Satellite_three_masses}
	\end{center}
\end{figure}

If $E$ is to decrease, then $a$ must become smaller; but if $H$ is constant, then $e$ must become smaller -- that is to say, that the orbit must round off. The quantity which remains constant is $a(1-e^{2})$, the \textit{semi-latus rectum} as shown in Figure \ref{fig:The_elliptical_orbit_of_a_satellite}. The planet is described by a point mass, $P$, and the satellite by a distribution of three masses, each $m/3$, at positions $S_{1},S_{2}$ and $S_{3}$, forming an equilateral triangle when free of stress. The masses are connected, as shown, by springs, each of unstressed	length $l$ and the same spring constant, $k$ (figure \ref{fig:Satellite_three_masses}). Thus a spring constantly stretched to a length $l'$ will exert an inward force equal to

\begin{equation} 
	F=k(l'-l)
\end{equation}

It is also important to introduce a dissipative element into the system by making the force dependent on the rate of expansion or contraction of the spring, giving the following force law:

\begin{equation} 
	F=k(l'-l)-c\frac{dl'}{dt} \label{eq:Formula 1.32}
\end{equation}
		
\noindent where the force acts inwards at the two ends. It is the second term in equation \ref{eq:Formula 1.32} which gives the simulation of the hysteresis losses in the satellite \cite[18-19]{Woolfson1999}.\\

This is a model for a computer simulation of a satellite in orbit around a planet subject to tidal stress. The satellite stretches along the radius vector in a periodic fashion, provided that the orbit is non-circular. Given that the satellite is not perfectly elastic, there will be hysteresis effects and some of the mechanical energy will be converted into heat and radiated away. For all purposes, nevertheless, the simulation fully specifies the target system.

Equations \ref{eq:Formula 1.31a} through \ref{eq:Formula 1.32} are a general description of the target system. Since the intention is to simulate a specific real-world phenomenon with concrete features, it must be singled out by setting the values for the parameters of the simulation. In the case of Woolfson and Pert, they use the following set of parameter values \cite[20]{Woolfson1999}:

\begin{enumerate}
	\item number of bodies = $4$
	\item mass of the first body (planet) = $2$ x $10^{27}$ kg
	\item mass of satellite = $3$ x $10^{22}$ kg
	\item initial time step = $10$ s
	\item total simulation time = $125000$ s
	\item body chosen as origin = $1$
	\item tolerance = $100$ m
	\item initial distance of satellite = $1$ x $10^{8}$ m
	\item unstretched length of spring = $1$ x $10^{6}$ m
	\item initial eccentricity = $0.6$
\end{enumerate}

These parameters set a computer simulation of a satellite of the size of Triton, the largest moon of Neptune orbiting around a planet with a mass close to Jupiter's -- including, of course, a specific tidal stress, hysteresis effects, and so forth. If the parameters were changed, then naturally the simulation is of another phenomenon -- though still of a two-body interaction using Newtonian mechanics. 

\begin{figure}
	\begin{center}
		\includegraphics[scale=0.3]{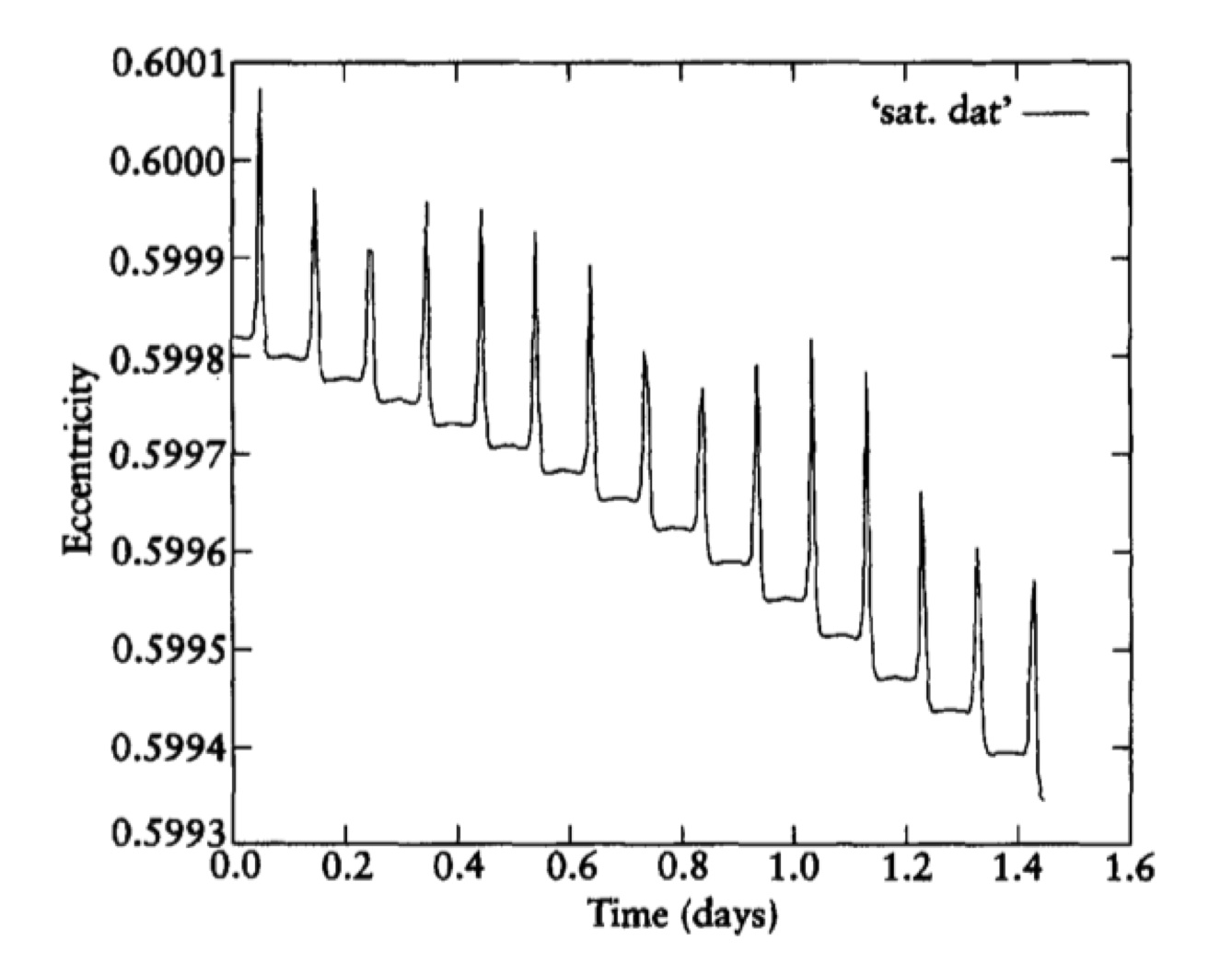}
		\caption{The orbital eccentricity as a function of time \cite[20]{Woolfson1999} \label{Fig3.3 Spikes}}
		\label{fig:Orbital_eccentricity}
	\end{center}
\end{figure}

 Here is an extract of the code corresponding to the above mathematical models as programmed in FORTRAN by Woolfson and Pert.\footnote{For the full code, see \cite{Woolfson1999a}.}

\tiny
\begin{verbatim} 

	  PROGRAM NBODY
       
				[...]
        
  C  THE VALUES OF A AND E ARE CALCULATED EVERY 100 STEPS 
  C  AND ARE STORED TOGETHER WITH THE TIME.
  C
      
      IST=IST+1
      IF((IST/100)*100.NE.IST)GOTO 50
      IG=IST/100
      IF(IG.GT.1000)GOTO 50
      
  C
  C  FIRST FIND POSITION AND VELOCITY OF CENTRE OF MASS OF THE 
  C  SATELLITE.
  C
      DO 1 K=1,3
      POS(K)=0
      VEL(K)=0
      DO 2 J=2,NB
      POS(K)=POS(K)+X(J,K)
      VEL(K)=VEL(K)+V(J,K)
      2 CONTINUE
      POS(K)=POS(K)/(NB-1.0)
      VEL(K)=VEL(K)/(NB-1.0)
      1 CONTINUE
       
  C
  C  CALCULATE ORBITAL DISTANCE
  C
      R=SQRT(POS(1)**2+POS(2)**2+POS(3)**2)

  C
  C  CALCULATE V**2
  C
      V2=VEL(1)**2+VEL(2)**2+VEL(3)**2
  C
  C  CALCULATE INTRINSIC ENERGY
  C
	  TOTM=CM(1)+CM(2)+CM(3)+CM(4)
      EN=-G*TOTM/R+0.5*V2      
				
				[...]        
        
      SUBROUTINE START
      DIMENSION X(20,3),V(20,3),ASTORE(1000),TSTORE(1000),
      +ESTORE(1000),XTEMP(2,20,3),VTEMP(2,20,3),CM(20),XT(20,3),
      +VT(20,3),DELV(20,3)
				[...]
      READ(5,*)NORIG

				[...]
  
  C
  C  INITIAL VELOCITIES ARE CALCULATED FOR THE THREE COMPONENTS OF THE 
  C  SATELLITE SO THAT THE SPIN ANGULAR VELOCITY OF THE SATELLITE IS
  C  APPROXIMATELY EQUAL TO THE ORBITAL ANGULAR VELOCITY.
  C 
	  VV=SQRT(G*(CM(1)+CM(2)+CM(3)+CM(4))*(1-ECC)/D)
	  V(2,2)=VV*DIS2/D
	  V(3,2)=VV*DIS3*COS(ANGLE)/D
				[...]
				
	RETURN
	END
\end{verbatim}

\normalsize

As discussed earlier, one of the characteristics of the problem-solving viewpoint is that the above mathematical model can be directly implemented on the computer. That is to say, that the code shown here corresponds in full extent to the mathematical model. In principle, nothing -- relevant -- is added and nothing -- relevant -- is eliminated. Thus, the only reason to use a computer is to find the set of solutions of the model in ways that are faster and cheaper. The description of patterns of behavior viewpoint, on the other hand, acknowledges the existence of a process for implementing the model as a computer simulation, which includes to turn the mathematical model into a rather different kind of model. 

As we can see, both positions hold good grounds. The problem-solving viewpoint is correct in claiming that computer simulations must reflect the mathematical model implemented, otherwise problems of representation, reliability, and the like emerge. The description of patterns of behavior viewpoint, on the other hand, reflects scientific and engineering practice more unambiguously.

Before continuing, here is a good place to introduce some new terminology. Let us call `mathematical models' those models used in scientific and engineering fields that make use of mathematical language. Examples of these are the above equations. Let us call `simulation models' those models implemented on a computer -- as a computer simulation -- which make use of a programming language.\footnote{Strictly speaking, a simulation model is a more complex structure consisting, among other things, of a \textit{specification} coded in a programming language as an \textit{algorithm} and finally implemented as a \textit{computer process}. Although the same specification can be written in different programing languages and implemented by different computer architectures, they are all considered the same simulation model. Thus understood, the programming language by itself does not determine the notion of `simulation model.' I shall discuss these issues in more extent in chapter \ref{chap:Units_of_analysis}.} An example of a simulation model is the code shown above. \\

I began this chapter making a distinction between two viewpoints of computer simulations. On the one hand, the problem-solving viewpoint which emphasizes the computational side of simulations; on the other hand, the description of patterns of behavior viewpoint which places emphasis on the representations of target systems. As mentioned before, the problem-solving viewpoint does not neglect the representation of the target system, nor does the description of patterns of behavior viewpoint fail to consider computation as a core issue of computer simulations. Neither viewpoint reflect an `all or nothing' stand -- a good example of this is the definition given by Thomas H. Naylor, Donald S. Burdick, and W. Earl Sasser where they defend a patterns of behavior viewpoint on page 1361 of \cite{Naylor1967} and also subscribe to the problem-solving viewpoint later on page 1319. The difference between these two viewpoints lies, again, in the main features highlighted by each account. Let us see if we can make this distinction more clear.\footnote{An interesting introduction to the history of computer science can be found in the work of \cite{Ceruzzi1998, DeMol2014, DeMol2015}, and particularly on computer simulations \cite{Oeren2011, Oeren2011a}.}

\subsection{Computer simulations as problem-solving techniques} \label{PS}

Under the problem-solving viewpoint, computer simulations typically exhibit some of the following features. First, simulations are adopted for cases where the target system is too complex to be analyzed on its own -- call it the \textit{complexity feature}. Second, simulations are useful for cases where the underlying mathematical model cannot be analytically solved -- call it the \textit{unanalyticity feature}. Third, mathematical models are directly implemented on the computer -- call it the \textit{direct implementation feature}. The complexity and unanalyticity features emphasize our human limitations for analyzing certain kind of mathematical models, at the same time enhancing the computational power simulations as a virtue. The direct implementation feature accompanies these ideas by claiming that there is no mediating methodology between the mathematical model and the physical computer. Rather, equations from the mathematical model are implemented -- or solved -- \textit{simpliciter} on the physical computer in the form of a computer simulation.

The early literature of the problem-solving viewpoint presents a rather uniform perspective on the matter. For the most part, professional philosophers, scientists, and engineers see the computational power of simulations as the key unlocking their epistemic power. A good first example is the definition provided by Claude McMillan and Richard Gonz\'{a}les in 1965. In their work, the authors state four characteristic points of simulations, namely

\begin{quotation}
	\begin{enumerate}
		\item Simulation is a problem solving technique.
		\item It is an experimental method.
		\item Application of simulation is indicated in the solution of problems of (a) systems design (b) systems analysis.
		\item Simulation is resorted to when the systems under consideration cannot be analyzed using direct or formal analytical methods. \cite{McMillan1965}
	\end{enumerate}
\end{quotation}

This definition is, to the extent that I could find, the first that openly conceives computer simulations as problem-solving techniques. This is not only because the authors explicitly state so in their first point, but because the definition adopts two of the three standard feature of this viewpoint. Point 4 is explicit about the use of simulations for finding solutions to otherwise unsolvable mathematical models, while point 3 suggests the adoption of simulations for system design and system analysis as they are too complex to be analyzed by their own (i.e., the complexity feature).

A year later, Daniel Teichroew and John Francis Lubin presented their own definition. Interestingly, this definition makes three features of this viewpoint more visible than any other definition in the literature. The authors begin by identifying what they call `simulation problems,' that is, problems that are treated by simulation techniques -- we shall discuss next what these techniques are. A simulation problem is basically a mathematical problem with many variables, parameters, and functions that cannot be treated analytically (i.e., the complexity feature) and thus computer simulations are the only available resource to researchers (i.e., the unanalyticity feature). The third feature, direct implementation of a mathematical model, can be found in several places in the article. In fact, the authors categorize two kinds of models, namely, \textit{continuous change models} (i.e., those making use of Partial Differential Equations or Ordinary Differential Equations) and \textit{discrete change models} (i.e., those models where changes in the state of the system are discrete) \cite[724]{Teichroew1966}. To the authors, both kinds of models are implemented directly as a computer simulation. In the authors' own words,

\begin{quote}
	Simulation problems are characterized by being mathematically intractable and having resisted solution by analytic methods. The problems usually involve many variables, many parameters, functions which are not well-behaved mathematically, and random variables. Thus, simulation is a technique of last resort. Yet, much effort is now devoted to `computer simulation' because it is a technique that gives answers in spite of its difficulties, costs and time required. \cite[724]{Teichroew1966}
\end{quote}

There is a further interesting claim to highlight here. Let us notice that the authors make plain the sentiment that advocates of this viewpoint have with regards to computer simulations: they are a technique of last resort.\footnote{Regarding this last point, Prof. \"Oren has organized in 1982 a NATO Advanced Study Institute in Ottawa focused on addressing the context for the uses of computer simulations (personal communication). See for instance, the articles published in \cite{Oeren1982, Oeren1984}.} That is to say, the use of computer simulations is only justified when analytic methods are unavailable. But this is more an epistemological prejudice against computer simulations than an established truth. Recent work done by philosophers shows that, in many instances, researchers prefer computer simulations over analytic methods. This, of course, for the obvious cases when the target system is intractable -- as Teichroew and Lubin correctly point out -- and where analytic solutions are not available. Vincent Ardourel and Julie Jebeile argue that computer simulations might even be superior to analytical solutions for the purpose of making quantitative predictions. According to these authors, ``some analytical solutions make numerical applications difficult or impossible (...) analytical solutions are sometimes too sophisticated with respect to the problem at stake (...) [and] analytical methods do not offer a generic approach for solving equations like [computer simulations do]''\footnote{The authors identify `numerical methods' with `computer simulations' \cite[202]{Ardourel2017}. As I show next, these two concepts must remain separate. However, this does not represent an objection to their main claim.} \cite[203]{Ardourel2017}. 

Now, advocates of the problem-solving viewpoint are also present in contemporary literature. A largely objected definition -- which, despite the author's change of mind somehow managed to become a standard in the literature -- is Humphreys' \textit{working definition}: ``A computer simulation is any computer-implemented method for exploring the properties of mathematical models where analytic methods are unavailable'' \cite[501]{Humphreys1990}.

Here, Humphreys gives us two features of computer simulations as problem-solvers. These are, simulations are mathematical models implemented on a computer, and they are used when analytic methods are unavailable. So far, Humphreys is a classic advocate of the problem-solving technique. A closer look of the definition, however, shows Humphreys' worries also include the nature of computation. Here is why.

Earlier, I mentioned that Naylor, Burdick, and Sasser stated that computer simulations are \textit{numerical methods} implemented on the computer. Humphreys, instead, conceives computer simulations as \textit{computer-implemented} methods. The distinction is not otiose, since it says something about the nature of computing a model. In fact, Humphreys urges to keep three different notions separate: \textit{numerical mathematics}, \textit{numerical methods}, and \textit{numerical analysis}. Numerical mathematics is the branch of mathematics concerned with obtaining numerical values of the solutions to a given mathematical problem. Numerical methods, on the other hand, are numerical mathematics concerned with finding an approximate solution to the model. Finally, numerical analysis is the theoretical analysis of numerical methods and the computed solutions \cite[502]{Humphreys1990}. Numerical methods, by themselves, cannot directly be related to computer simulations. At least two additional features must be included. First, numerical methods must be applied to a specific scientific problem. This is important because the model implemented is not \textit{any} model, but of a specific kind (i.e., scientific and engineering models). In this way, there is no room for conflating computer simulations carried out in a scientific facility with computer simulation carried out for artistic purposes. Second, the method must be implemented on a real computer as well as computable in real time. This second feature ensures that the model is suitable for computation and complies with minimal standards of scientific research (e.g., that the computation ends within a reasonable time-frame, that the results are accurate within a certain range, etc.) 

Despite being suggested only as a working definition, Humphreys received fierce objections that virtually forced him to change his original position. A chief critic was Stephan Hartmann, who objected that Humphreys' definition missed the dynamic nature of computer simulations. Hartmann, then, offered his own definition:

\begin{quotation}
Simulations are closely related to dynamic models. More concretely, a simulation results when the equations of the underlying dynamic model are solved. This model is designed to imitate the time-evolution of a real system. To put it another way, \textit{a simulation imitates one process by another process}. In this definition, the term ``process" refers solely to some object or system whose state changes in time. If the simulation is run on a computer, it is called \emph{a computer simulation} \cite[83 - emphasis in original]{Hartmann1996}.
\end{quotation}

Simplifying this definition, we could say that a computer simulation consists of finding the set of solutions to a dynamic model by using a physical computer. Let us highlight a few interesting assumptions. First, the dynamic model is conceived as holding no differences from a mathematical model. Thus understood, the model used by M. M. Wolfson and G. J. Pert for simulating the dynamics of a satellite orbiting around a planet under tidal stress \textit{is} the model implemented on the physical computer. Second, Hartmann is not too worried about which methods are used for solving the dynamic model. Paper and pencil, numerical methods, and computer-implemented methods seem to be all equally suitable. This concern stems from taking that the same dynamic model is solved by a human agent as well as the computer. Similarly to Naylor, Burdick, and Sasser, such an assumption raises questions about the nature of computation.

It is interesting to note that Hartmann's definition has been warmly welcomed by the philosophical community. The same year, Jerry Banks, John Carson, and Barry Nelson presented a definition similar to Hartmann's, also emphasizing the idea of \emph{the dynamics of a process over time}, and of representation as \emph{imitation}. They define it in the following way ``[a] simulation is the imitation of the operation of a real-world process or system over time. Whether done by hand or on a computer, simulation involves the generation of an artificial history of a system and the observation of that artificial history to draw inferences concerning the operating characteristics of the real system'' \cite[3]{Banks2010}. Francesco Guala also follows Hartmann in distinguishing between \emph{static} and \emph{dynamic} models, time-evolution of a system, and the use of simulations for mathematically solving the implemented model \cite{Guala2002}. More recently, Wendy Parker has made explicit reference to it by characterizing a simulation as ``a time-ordered sequence of states that serves as a representation of some other time-ordered sequence of states'' \cite[486]{Parker2009}. 

Now, despite of the differences between Humphreys and Hartmann, they agree on a few issues as well. In fact, they both consider computer simulations as high-speed calculation equipment capable of enhancing our analytical capacity to solve otherwise unsolvable mathematical models. After Hartmann's initial objections, Humphreys coined a new definition, this time based on the notion of \emph{computational template}. I shall discuss templates in the next section, as I believe this new conceptualization of computer simulations qualifies better for the descriptions of patterns of behavior viewpoint.

An illuminating summary is found in the work of Roman Frigg and Julian Reiss. According to the authors, there are two senses in which the notion of \emph{computer simulation} is defined in current literature. There is a \textit{narrow sense}, where  ```simulation' refers to the use of a computer to solve an equation that we cannot solve analytically, or more generally to explore mathematical properties of equations where analytical methods fail''.  There is also a \textit{broad sense}, where the term ```simulation' refers to the entire process of constructing, using, and justifying a model that involves analytically intractable mathematics'' \cite[596]{Frigg2009a}. To me, both senses could be included as part of the problem-solving techniques viewpoint of computer simulations.

Both categories are certainly meritorious and illuminating. Both generally capture the many senses in which philosophers of the problem-solving viewpoint define the notion of \emph{computer simulation}. While the narrow sense focuses on the heuristic capacity of computer simulations, the broad sense emphasizes the methodological, epistemological, and pragmatic aspects of computer simulations as problem-solvers. Let us now move to a different way to conceptualize computer simulations.

\subsection{Computer simulations as description of patterns of behavior}

The view of computer simulations as problem solving techniques contrasts with the view of simulations as description of patterns of behavior. Under this view, computer simulations are primarily concerned with describing the behavior of a target system to which they develop or unfold. As mentioned before, this is not to say that the computing power of simulations is downplayed in any sense. Computer simulations as problem solvers got this point right in the sense that speed, memory, and control are core factors that emphasize the novelty of simulations in scientific and engineering practice.  However, under this viewpoint, the computational power of simulations is considered a second-level feature. In this sense, instead of locating the epistemological value of computer simulations in their capacity for solving a mathematical model, their value comes from describing patterns of behavior of target systems. 

Now, what are patterns? I take them to be descriptions that reflect structures, attributes, performance, and the general behavior of the target system in a specific language. More specifically, these structures, attributes, and so on are interpreted as concepts used in the sciences (e.g., H$_2$O, mass, etc.), causal relationships (e.g., the collision of two billiard balls), natural and logical necessities (e.g., that no enriched uranium sphere has a mass greater than 100,000 kilograms\footnote{A clarification is due here. A computer is not \textit{technologically} impaired to simulate an enriched uranium sphere with a mass greater than 100,000 kilograms. Rather, the kind of constraints we find in computers are related to their own physical limitations and those indicated by theories of computation. Now, given that researchers want to simulate a \textit{real} target system, they must describe it as accurately as possible. If that target system is a natural system, such as a uranium sphere, then accuracy dictates that the simulation is limited on the mass of the sphere.}), laws, principles, and constants of nature. In short, patterns are descriptions of a target system which make use of the scientific and engineering vocabulary. Naturally, these patterns also rely on expert knowledge, `tricks of the trade', past experiences, and individual, societal and institutional preferences. In this sense, for this viewpoint, computer simulations are a conglomerate of concepts, formulae, and interpretations that facilitate the description of the patterns of behavior of a target system.

The difference in conceptualizing computer simulations in this way, as opposed to the problem-solving viewpoint, is that the physical features of the computer are no longer the primary epistemic value of computer simulations. Rather, it is their capacity to describe patterns of behavior of a target system that carries the burden. Mimicking the previous section, let us begin with some early definitions. 

In 1960, Martin Shubik defined a simulation in the following way: 

\begin{quotation}
	A simulation of a system or an organism is the operation of a model or simulator which is a representation of the system or organism. (...) The operation of the model can be studied and, from it, properties concerning the behavior of the actual system or its subsystem can be inferred. \cite[909]{Shubik1960}
\end{quotation}

Shubik is highlighting two main features that are at the heart of this view. That is, that a simulation is a representation or description of the behavior of a target system, and that properties of such a target system can be inferred. The first feature is central to this viewpoint, to the extent that it gives the name to it. Emphasizing the representational capacity of simulations, as opposed to mathematical models, suggests that they are somehow different. As we shall see later, this difference lies in the number of transformations under which a mathematical model -- or rather, a series of mathematical models -- go through in order to result in a computer simulation. The second feature, on the other hand, highlights the use of computer simulations as proxies for understanding something about the target system. This is to say that researchers are capable of inferring properties of the target system based on the results of the simulation. 

Both features, we must notice, are absent in the problem-solving viewpoint. The contrary, however, is not true. As mentioned earlier, understanding computer simulations this way does not disavow some claims of the problem solving techniques viewpoint. In particular, the capacity of computing complex models is a characteristic of computer simulations usually present in all definitions. For instance, Shubik says: ``The model is amenable to manipulations which would be impossible, too expensive or impracticable to perform on the entity it portrays'' \cite[909]{Shubik1960}. It is interesting to note that, as we move forward in time, concerns about the computational power tend to disappear. 

Almost two decades later, in 1979, G. Birtwistle formulated the following definition for computer simulations:

\begin{quotation}
	Simulation is a technique for representing a dynamic system by a model in order to gain information about the underlying system. If the behaviour of the model correctly matches the relevant behaviour characteristics of the underlying system, we may draw inferences about the system from experiments with the model and thus spare ourselves any disasters. \cite[1]{Birtwistle1979}
\end{quotation}

Similarly to what Teichroew and Lubin presented with the problem-solving viewpoint, Birtwistle is also making plain the chief features of the description of patterns of behavior viewpoint. From the above definition it is clear that central to computer simulations is the representation of a target system; provided the right representation, then, researchers can draw inferences about such target system. Let us also note that, unlike Teichroew and Lubin who consider computer simulations a last resource, to Birtwistle it is a crucial piece in scientific research that helps to prevent disasters. The opposite attitudes towards computer simulations cannot find two better representatives.

Another definition worth mentioning comes from Robert E. Shannon, an industrial engineer who has worked very profusely on clarifying the nature of computer simulations (see his work from \cite{Shannon1975} and \cite{Shannon1978}).

\begin{quotation}
	We will define simulation as the process of designing a model of a real system and conducting experiments with this model for the purpose of understanding the behavior of the system and/or evaluating various strategies for the operation of the system. Thus it is critical that the model be designed in such a way that the model behavior mimics the response behavior of the real system to events that take place over time. \cite[7]{Shannon1998}
\end{quotation}

Again we can see how Shannon highlights the importance of representing a target system, as well as the ability to infer -- and evaluate -- our knowledge from computer simulations. What is perhaps the most outstanding aspect of Shannon's definition is the marked emphasis on the methodology of computer simulations. To him, it is critical that the model in the simulation mimics the behavior of the target system. It is not enough, as it is found in other authors, that the model correctly describes the relevant behavior of the target system. Attention must be given to the way in which the simulation is designed, because it is there where we will find grounds -- and problems -- for drawing inferences about the target system. 

These latter ideas continue, more or less successfully, in the subsequent literature related to this viewpoint. A good example is Paul Humphreys' 2004 book, where he presents a detailed account of the methodology of computer simulations. Eric Winsberg, a few years later, also made an interesting effort to show the ways in which design decisions affect epistemological evaluations. According to Winsberg, present and past design decisions ground our confidence in the results of computer simulations. Let us now discuss their positions in more detail.\footnote{There are many other contemporary authors that deserves our attention. Most prominently is the work of Claus Beisbart, who takes computer simulations as \textit{arguments} \cite{Beisbart2012}. That is, an inferential structure encompassing a premise and a conclusion. Another interesting case is Rawad El Skaf and Cyrille Imbert \cite{ElSkaf2013}, who conceptualize computer simulations as `unfolding scenarios.' Unfortunately, space does not allow me to discuss these authors in more extent.}

Earlier, I mentioned that in 1990 Humphreys elaborated on a \textit{working definition} for computer simulations. In spite of having presented it only as a working definition, he received vigorous objections that virtually forced him to change his original viewpoint on computer simulations. One of the chief critics was Stephan Hartmann, who pointed out that his working definition missed the dynamic nature of computer simulations.  After Hartmann’s initial objections, Humphreys coined a new definition, this time based on the notion of \textit{computational model}. 

According to his new characterization, computer simulations rely on an underlying computational model that involves representations of a target system. At first glance, this definition looks very much like the standard definitions discussed so far. However, the devil is in the details. In order to fully appreciate Humphreys' turn, we must dissect his definition of \textit{computational model}, understood as the sextuple:\\

\textit{$<$computational template, construction assumptions, correction set,\\
\indent interpretation, initial justification, output representation$>$}\footnote{For details, see \cite[102-103]{Humphreys2004}}\\

A \textit{computational template} is, in fact, the result of a computationally tractable theoretical template. A \emph{theoretical template}, in turn, is the kind of very general mathematical description that can be found in a scientific work. This includes partial differential equations, such as elliptic (e.g., Laplace's equation), parabolic (e.g., the diffusion equation), hyperbolic (e.g., the wave equation), and ordinary differential equations, among others. An illuminating example of a theoretical template is Newton's Second Law, for it describes a very general constraint on the relationship between forces, mass, and acceleration. The core characteristic of theoretical templates is that researchers could specify them in a number of different ways. For instance, the force function in Newton's Second Law could either be a gravitational force, an electrostatic force, a magnetic force, or any other kind of force. 

Now, a computational template cannot simply be picked out from the theoretical template. This is the kind of feature that drives the solving-problem viewpoint, but not the description of patterns of behavior viewpoint. To the latter viewpoint, there is an entire methodology that mediates between the computational model and the theoretical model that needs to be explored. Concretely, the process of construction of a template involves a number of idealizations, abstractions, constraints, and approximations of the target system for which researchers needs to account. Moreover, at some point the computational template needs to be validated against data. What happens when it fails to fit those data? Well, the answer is that researchers have a series of well-established methods for correcting the computational template in order to ensure accurate results. According to Humphreys, the \textit{construction assumptions} and \textit{correction set} -- components two and three in the sextuple -- fulfill precisely these roles. Without them, the computational template might not even be computable.

Now, in order to have an accurate representation of the target system, the variables, functions, and the like in the computational template need to be given an \textit{interpretation}. For example, in the first derivation of a diffusion equation, the interpretation of the function representing the temperature gradient in a perfectly insulated cylindrical conductor is central to the decision on whether the diffusion equation correctly represents the flow of heat in a given metal bar \cite[80]{Humphreys2004}. The researcher's interpretation of the computational template constitute part of the justification for adopting certain equations, values, and functions. The computational templates, says Humphreys, are ``not mere conjectures but objects for which a separate justification for each idealization, approximation, and physical principle is often available, and those justifications transfer to the use of the template.'' \cite[81]{Humphreys2004}.

Finally, the output representation, that is, the visualization of computational model, comes in different flavors. It can be a data array, functions, matrix, and more important in terms of understanding, dynamic representations such as videos or interactive visualizations. As we will discuss in detail in section \ref{Visualization_of_CS_results}, visualizations play a fundamental role in our epistemic gain using computer simulations, and thus in their general success as novel methods in scientific and engineering research.

Eric Winsberg is the second philosopher on our list. According to him, there are two core characteristics that meaningfully distinguish computer simulations from other forms of calculation. First, much effort goes into setting up the model that serve as the basis for computer simulations, as well as to deciding which simulation results are reliable and which are not. Second, computer simulations make use of a variety of techniques and methods that facilitate drawing inference from results \cite{Winsberg2010}. As discussed earlier in this section, these two characteristics are typical from the description of patterns of behavior viewpoint. 

Furthermore, Winsberg correctly points out that the construction of computer simulations are guided, but not determined, by theory. This means that, although computer simulations rely on theoretical background, they typically encompass elements that are not directly related to, nor are part of, theories. A case of this are `fictionalizations,' that is, \textit{contrary-to-fact} principles that are included in the simulation model with the purpose of increasing the reliability and trustworthiness of its results. As we saw earlier, Humphreys made a similar point with the construction assumptions and correction set. Winsberg then illustrates fictionalizations with two examples, `artificial viscosity' and `vorticity confinement.' In simulations of fluid dynamics, these techniques are successfully used despite not offering realistic accounts of the nature of fluids. Why are they used then? There are several reasons, including of course that they are largely part of the practice of model-building techniques on fluid dynamics. Other reasons include the fact that these fictionalizations facilitate the calculation of crucial effects that would otherwise be lost, and that without these fictionalizations, the results of simulations on fluid dynamics could neither be accurate nor justified.\\

The previous discussions show that it is simply not possible to fit the concept of computer simulations into one conceptual corset. Thus, our initial question: `what are computer simulations?' cannot be uniquely answered. It seems that, ultimately, it will depend on the commitments of the practitioners. Whereas the problem-solving viewpoint is more interested in finding solution to complex models, the description of patterns of behavior viewpoint is concerned with accurately representing a target system. Both offer good conceptualizations of computer simulations, and both have several problems to face. Let us next discuss three different kinds of computer simulations found in the scientific and engineering practice.

\subsection{Kinds of computer simulations \label{sec:Classes-of-CS}}

Before addressing the different classes of computer simulations, let us briefly discuss a short classification of target systems typically associated with computer simulations. This classification, besides being non-exhaustive -- or precisely because of this -- holds no expectations of being unique. Other ways of characterizing target systems -- along with the models that represent such target systems -- can lead to a new and improved taxonomy.

Having mentioned all the usual warnings, let us begin with the most familiar of all target systems, that is, \emph{empirical target systems}. These are empirical phenomena -- or real-world phenomena -- in all forms and flavors. Examples include microwave background radiation and Brownian movement in astronomy and physics, social segregation in sociology, competition among vendors in economy, and scramjets in engineering, among many others examples. 

Understandably, empirical target systems are the most pervasive target system in computer simulation. This is chiefly because researchers are seriously engaged in understanding the empirical world, and computer simulations provide a new and successful method for achieving such aims. Now, in order to represent empirical target systems, computer simulations implement models that theoretically underpin real-world phenomena with the help of laws, principles, and theories accepted by the scientific community. The Newtonian model of planetary movement, for instance, describes the behavior of any two bodies interacting with each other by a handful of laws and principles. Unfortunately, not every empirical target system can be so simply and accurately represented.

More commonly, computer simulations represent real-world phenomena by including a plethora of elements from different -- and sometimes incompatible -- sources. Take for instance \textit{scramjets}, combustion ramjets in which combustion takes place in supersonic airflow. The use of Navier-Stoke equations are typically at the base of simulations of fluid dynamics. However, the intake of a scramjet compresses the incoming air via a series of shock waves generated by the specific shape of the intake along with the high flight velocity, as opposed to other air-breathing vehicles that compress the incoming air by compressors -- or other moving parts. Simulations of scramjets, then, cannot be fully characterized by Navier-Stoke equations. Instead, the laminar and turbulent boundary layers, along with the interaction with shock waves, yields a three-dimensional unsteady complex flow pattern. A reliable simulation of what is happening within the intake is, then, accomplished by means of high fidelity direct numerical simulations and large eddy simulations. It is the model design, programmed, and built by the engineers, and not just the Navier-Stoke equations, that permits a reliable simulation \cite{Barnstorff2010}.\footnote{I should also mention that there are several other contrivances also involved in the design and programming of computer simulations. In this respect, section \ref{Trust_on_CS} presents and discusses some of them, such as calibration procedures, and verification and validation methods.}

Another important target system is the so-called \emph{hypothetical target system}. These are target systems where no empirical phenomena are described. Rather, they are either \textit{theoretical} or \textit{imaginary}.  A \textit{theoretical target system} describes systems or processes within the universe provided by a theory, whether mathematical (e.g., a torus), physical (e.g., air resistance equal to zero), or biological (e.g., infinite populations). Take as example the famous problem of the Seven Bridges of K\"{o}nigsberg,\footnote{The problem can be best described as finding a way to cross each of the seven bridges of the city of K\"onigsberg only once. The problem, solved by Euler in 1735, laid the foundations of graph theory.} or the Traveling Salesman Problem.\footnote{The Travelling Salesman Problem describes a salesman who must travel between \textit{N} number of cities and keep the travel costs as low as possible. The problem consists in finding the best optimization of the the salesman's route.} Thus understood, the target system is not empirical, but instead it has the properties of a mathematical or a logical system. A computer simulation implementing these models is mostly theoretical in essence, and it is typically designed for exploring the underlying properties of the model.

\textit{Imaginary target systems}, on the other hand, stand for non-existing, imagined scenarios. For instance, an epidemic outbreak of influenza in Europe counts as such a target system. This is because such a scenario is prone to never exist, although it does not mean that it will never happen. A simulation of such a scenario provides researchers the necessary understanding of the dynamics of an epidemic outbreak for planning prevention measures and containment protocols, as well as for training personnel. Imaginary systems can be, in turn, divisible into two further kinds, namely, \textit{contingent} and \textit{impossible} \cite{Weisberg2013}. The former stands for a scenario that, as a matter of contingent fact, does not exist. The latter stands for a scenario that is nomologically impossible. The simulation of an epidemic outbreak is a case in point of the former, while running a simulation that violates the known laws of nature is an example of the latter.

The first thing to note about this classification is that computer simulations could represent one target system but render results of another target system. This is a common `jumping' mechanism that could be harmless, or that could cast a shadow of doubt on the results. A simple example will show how this is possible. Consider simulating the planetary movement implementing a Newtonian model; now instantiate $G = 2m^3 kg^{-1} s^{-2}$ as an initial condition. The example shows a simulation that initially implements an \textit{empirical target system}, but renders results of a \textit{nomologically impossible imaginary target system}. To the best of our current knowledge of the universe, there is no such gravitational constant. As a consequence, the results of a simulation which in principle should have been sanctioned empirically (e.g., by validating against empirical data) can only be confirmed theoretically.\footnote{Running a second computer simulation that could confirm these results is becoming standard practice \cite{Ajelli2010}.}  

In a way, these issues are part of the general charm and malleability of using computer simulations, but they need to be taken seriously by philosophers and sociologists science. Having said this, and looking closely at the practice of computer simulation, one could sees how researchers have a few `tricks' that help to cope with situations like `jumping'. For instance, one solution to the simulation of the satellite under tidal stress would be to set $G$ as a global constant of value $6.67384x10^{-11} m^3 kg^{-1} s^{-2}$. Unfortunately, this is only a palliative solution since the value of the variable \textit{mass of satellite} could be set to any unrealistically large value. Again, researchers could establish lower and upper limits on the size of the satellite and planet mass, but this solution only begs the question if there is not another way to `fool the simulation'. 

Either by modeling or by instantiation, computer simulations can create several scenarios out of the mind of researchers. How do researchers turn this seemingly disastrous situation into something advantageous? The answer is, I believe, in the way scientists and engineers sanction computer simulations as reliable processes. That is, by providing reasons for believing that computer simulations are a reliable process which renders, most of the time, correct results. I will explore these issues in more detail in chapter \ref{Knowledge_from_CS}. 

Along with a classification of target systems, I now suggest a classification of computer simulations. In the same spirit, this classification is not meant to be exhaustive, conclusive, nor unique. Here, I divide simulations into three classes, based on the standard treatment that computer simulations have received from the specialized literature \cite{Winsberg2015}). These are \textit{cellular automaton}, \textit{agent-based simulations}, and \textit{equation-based simulations}. 

\subsection{Cellular automata}

Cellular automata are the first of our examples of computer simulations. They were devised in the 1940s by Stanislaw Ulam and John von Neumann while Ulam was studying the growth of crystals using a simple lattice network as a model, and von Neumann was working on the problem of self-replicating systems. The story goes that Ulam suggested to von Neumann to use the same kind of lattice network as his, creating in this way a two-dimensional, self-replicator algorithm. 

Cellular automata are simple forms of computer simulations. Such simplicity stems from both their programming and underlying conceptualization. A standard cellular automaton is an abstract mathematical system where space and time are considered to be discrete; it consists of a regular grid of cells, each of which can be in any state at a given time. Typically all the cells are governed by the same rule, which describes how the state of a cell at a given time is determined by the states of itself and its neighbors at the preceding moment. Stephen Wolfram defines cellular automata in the following way:

\begin{quote}
	
	[...] mathematical models for complex natural systems containing large numbers of simple identical components with local interactions. They consist of a lattice of sites, each with a finite set of possible values. The value of the sites evolve synchronously in discrete time steps according to identical rules. The value of a particular site is determined by the previous values of a neighborhood of sites around it. \cite[1]{Wolfram1984}
	
\end{quote}

Although a rather general characterization of this class of computer simulation, the above definition already provides the first ideas as to their domain of applicability. Cellular automata have been successfully used for modeling many areas in social dynamics (e.g., behavioral dynamics for cooperative activities), biology (e.g., patterns of some seashells), and chemical types (e.g., the Belousov-Zhabotinsky reaction). 

One of the simplest and most canonical examples of cellular automata is Conway's \textit{Game of Life}. The simulation is remarkable because it provides a case of emergence of patterns and self-organization dynamics of some systems. In this simulation, a cell can only survive if there are either two or three other living cells in its immediate neighborhood. Without these companions, the rule indicates that the cell dies either from overcrowding, if it has too many living neighbors, or from loneliness, if it has too few. A dead cell can make its way back to life provided that there are exactly three living neighbors. In truth, there is little interaction -- as one would expect from a game -- besides creating an initial configuration and observing how it evolves. Nevertheless, from a theoretical point of view, the Game of Life can compute any computable algorithm, making it a remarkable example of a universal Turing machine. Back in 1970, Conway's Game of Life open up a new field of mathematical research: the field of cellular automata \cite{Gardner1970}. 

\begin{figure}

	\begin{center}
		\includegraphics[scale=0.50]{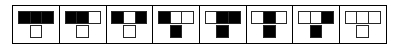}
		\caption{Rule 30. http://mathworld.wolfram.com/CellularAutomaton.html}
		\label{fig:Rule_30}	
	\end{center}
	
\end{figure}

\begin{figure}
	\begin{center}
		\includegraphics[scale=0.30]{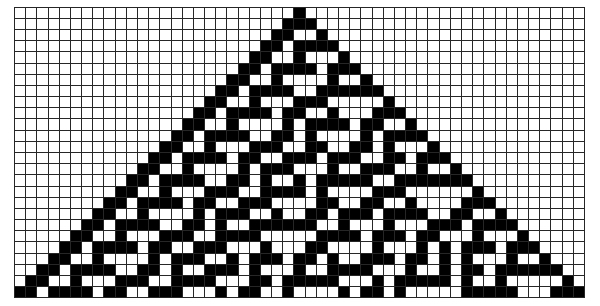}
		\caption{Pattern created by Rule 30. http://mathworld.wolfram.com/CellularAutomaton.html}
		\label{fig:Rule_30_Automaton}
	\end{center}
\end{figure}

Elementary cellular automata furnishes some fascinating cases in contemporary science. The idea of these automata is that they are based on an infinite one-dimensional array of cells with only two states. At discrete time intervals, every cell changes state based on its current state and the state of its two neighbors. Rule 30 is a case in point which produces complex, seemingly random patterns from simple, well-defined rules (see figure \ref{fig:Rule_30}). For instance, a pattern resembling Rule 30 appears on the shell of cone snail species \textit{Conus textile} (see figure \ref{fig:Rule_30_Automaton}). Other examples are based on it's mathematical properties, such as using Rule 30 as a random number generator for programming languages, and as a possible stream cipher for use in cryptography. The rule set which governs the next state of Rule 30 is shown in figure \ref{fig:Rule_30}.

Cellular automata entrench a set of unique methodological and epistemological virtues. To name a few, they accommodate better to error because they render exact results of the model they implement. Since approximations with the target system are almost nonexistent, any disagreement between the model and the empirical data can be ascribed directly to the model that realized the set of rules. Another epistemological virtue is pointed out by Evelyn Fox-Keller, who explains that cellular automata lack theoretical underpinning in the familiar sense of the term. That is, ``what is to be simulated is neither a well-established set of differential equations [...] nor the fundamental physical constituents (or particles) of the system [...] but rather the phenomenon itself'' \cite[208]{Keller2003}. Approximations, idealizations, abstractions and the like are concepts that worry the practitioner of cellular automata very little.

Now, not everything is great for cellular automata. They have been criticized on several grounds. One of these criticisms touches upon the metaphysical assumptions behind this class of simulation. It is not clear, for instance, that the natural world is actually a discrete place, as assumed by the cellular automata. Many of today's scientific and engineering work is based on the description of a continuous world. On less speculative grounds, it is a fact that cellular automata have little presence in scientific and engineering fields. The reason for this, I believe, is partly cultural. The physical sciences are still the accepted viewpoint for describing the natural world, and they are written in the language of Partial Differential Equations and Ordinary Differential Equations (PDE and ODE respectively). 

Naturally, advocates of cellular automata focus their efforts to show their relevance. In all fairness, many cellular automata are more adaptable and structurally similar to empirical phenomena than PDEs and ODEs \cite[vii]{Wolfram1984a}. It has been pointed out by Annick Lesne, a renown theoretical physicist, that discrete and continuous behavior coexist in many natural phenomena, depending on the scale of observation. To her mind, this is an indicator not only of the metaphysical basis of many natural phenomena, but also of the suitability of cellular automata for scientific and engineering research \cite{Lesne2007}. In a similar vein, G\'erard Vichniac believes that cellular automata seek not only numerical agreement with a physical system, but also they attempt to match the simulated system’s own structure, its topology, its symmetries and its `deep' properties \cite[113]{Vichniac1984}. Tommaso Toffoli has a similar stand as these authors, to the point that he entitled a paper: ``Cellular automata as an alternative to (rather than an approximation of) differential equations in modeling physics'' \cite{Toffoli1984}, highlighting cellular automata as the natural replacement of differential equations in physics. 

Despite these and many other authors' efforts to show that the world might be more adequately described by cellular automata, the majority of scientific and engineering disciplines have not made a complete shift yet. Most of the work done in these disciplines is predominantly based on agent-based and equation-based simulations. As mentioned before, in the natural sciences and engineering, most physical and chemical theories used in astrophysics, geology, climate change, and the like implement PDE and ODE, two systems of equations that are at the basis of equation-based simulations. Social and economic systems, on the other hand, are better described and understood by means of agent-based simulations.

\subsection{Agent-based simulations}

While there is no general agreement on the definition of the nature of an `agent', the term typically refers to self-contained programs that control their own actions based on the perceptions of the operating environment. In other words, agent-based simulations `intelligently' interact with their peers as well as their environment. 

The relevant characteristic of these simulations is that they show how the total behavior of a system emerges from the collective interaction of their parts. To deconstruct these simulations into their constituent elements would remove the added value that has been provided in the first place by the computation of the agents. It is a fundamental characteristic of these simulations, then, that the interplay of the various agents and the environment brings about a unique behavior of the entire system. 

\def\Schelling{Although nowadays Schelling's model is carried out by computers, Schelling himself warned against its use for understanding the model. Instead, he used coins or other elements to show how segregation occurred. In this respect, Schelling says: ``I cannot too strongly urge you to get the nickels and pennies and do it yourself. I can show you an outcome or two. A computer can do it for you a hundred times, testing variations in neighborhood demands, overall ratios, sizes of neighborhoods, and so forth. But there is nothing like tracing it through for yourself and seeing the process work itself out. It takes about five minutes – no more time than it takes me to describe the result you would get'' \cite[85]{Schelling1971}. Schelling's warning against the use of computers is an amusing anecdote that illustrates how scientists could sometimes fail in predicting the role of computers in their own respective fields.}

\begin{figure}
	\begin{center}
			\includegraphics[scale=1.0]{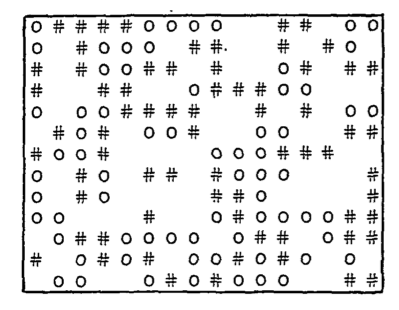}
			\caption{Initial random distribution of agents in a checkerboard 13 rows $x$ 16 columns, with a total of 208 squares.
					\label{fig:Schelling_matrix}	
					 \cite[155]{Schelling1971}}
	\end{center}
\end{figure}

Good examples of agent-based simulations proceed from the social and behavioral sciences, where they are heavily present. Perhaps the most well-known example of an agent-based simulation is Schelling's Model of Social Segregation.\footnote{\Schelling} A very simple description of Schelling's model consists in two groups of agents living in a 2-D\footnote{Schelling also introduced a 1-D version, with a population of 70 agents, with the four nearest neighbors on either side, the preference consists of not being minority, and the migration rule is that whoever is discontent moves to the nearest point that meets her demands \cite[149]{Schelling1971}.}, \textit{n x m} matrix  `checkerboard' where agents are placed randomly. Each individual agent has a \textit{3 x 3} neighborhood, which is evaluated by a utility function that indicates the migration criteria. That is, the set of rules that indicates how to relocate -- if possible -- in case of discontent by an agent (see Fig. \ref{fig:Schelling_matrix})

Schelling's segregation model is a canonical example of agent-based simulations.\footnote{In truth, depending on how the Schelling model is design and programmed, it could also qualify as an Cellular Automata. Thank you Andr\'es Il\v{c}i\'c for pointing this out to me.} But more complex agent-based simulations can be found in the literature. It is now standard that researchers model different attributes, preferences, and overall behavior into the agents. Nigel Gilbert and Klaus Troitzsch list the set of attributes that are typically modeled by agents \cite{Gilbert2005}: 

\begin{enumerate}
	\item \textit{Knowledge and belief}: Since agents base their actions on their interaction with their environment as well as other agents, it is crucial to be able to model their system of beliefs. The traditional distinction between \textit{knowledge}, as true justified belief, and mere \textit{belief} can then be modeled. Those whose information might be faulty or plain false must be modeled to act in their environment in a rather different way than those agents whose information is correct, as a result of knowing.
	
	\item \textit{Inference}: Knowledge and belief are possible because agents are able to infer information from their own set of beliefs. Such inferences are modeled in different ways, sometimes over very intuitive assumptions. For instance, agent \textit{A} could infer that a source of `food' is near agent \textit{B} knowing -- or simply believing -- that agent \textit{B} has `eaten some food.' 	
	
	\item \textit{Goals}: Since agents are, for the most part, programed as autonomous entities, they are typically driven by some sort of goals. Survival is a good example of this, as it might require the satisfaction of subsidiary goals, such as acquiring sources of energy, food, and water, as well as avoiding any dangers. Modeling these subsidiary goals is not easy, since researchers must decide on how to weigh the importance and relevance among several goals. Different design decisions lead to different goal-guided agents, and therefore to a different overall behavior of the system. Let us note that the goals of an agent are a different attribute from knowing and inferring. Whereas the former are about guiding the general behavior of the agents in their environment, the latter depends on valid or true information for ruling their behavior. 
	
	\item \textit{Planning}: In order to satisfy its goals, an agent needs to have some way of determining what decisions are best. Typically, a set of condition-action rules are programed as constituent of the agent. For instance, a utility function is programed for the satisfaction of an energy source and for what counts as a `danger.' Planning involves inferring what actions lead to a desired goal, what state is required before that action takes place, and what actions are needed to arrive at the desired state. In this respect, planning is very sophisticated as the agent needs to weigh several options, including having a `pay-off' rule over their decisions, determining where it needs to be at some point in the future, and the like. It has been objected that agent planning is not realistic of human planning because most human action is driven by routine decisions, an inherent tendency to discern, and even instinctive judgments that cannot be modeled by a calculated plan. 

	\item \textit{Language}: Passing information across agents is central for any agent-based simulation. An exciting example is Alexis Drogoul and Jacques Ferber's multi-agent model of ants' colonies \cite{Drogoul1992}. According to the authors, agents can communicate by propagating a `stimuli' to the environment. This stimuli might be received and transmitted in different ways. When ants receive this stimuli, they activate a friendly behavior. However, when a predator receives the stimuli, it triggers an aggressive behavior.	In this particular agent-based simulation, such a communication mechanism is very simple and by no means aims at conveying any meaning. 	Other examples of modeling language in agent-based simulations are the negotiation of contracts \cite{Smith1981}, communication of decisions, and even one agent threatening another with `death' \cite{Gilbert2005}.
	
	\item \textit{Social models}: Some agent-based simulations, like Schelling's segregation model discussed earlier, aim at modeling the interrelationship between agents in a larger environment. On this account, agents are able to create their own topology given the set of rules, their interaction with other agents, and the initial setup, among other parameters. For instance, in Schelling's segregation model, the agents create different topologies of segregation given the side of the grid, the utility function, and the initial position of the agents. 
\end{enumerate}

\subsection{Equation-based simulations} \label{subsec:Equation-based_simulation}

Perhaps the most commonly used class of computer simulations in science and engineering are the so-called equation-based simulations. At their basis, they are the implementation of a mathematical model on the physical computer which describes a target system. Because most of our understanding of the world comes through the use of mathematical descriptions, these simulations are by far the most popular in scientific and engineering fields. Naturally, examples abound. Fluid dynamics, solid mechanics, structural dynamic, shock wave physics, and molecular chemistry, are just a handful of cases. William Oberkampf, Timothy G. Trucano, and Charles Hirsch have elaborated on an extensive list that they deem to call `computational engineering' and `computational physics' \cite{Oberkampf2003}. Let us note that their labeling emphasizes simulations \textit{only} in physics and engineering fields. Although it is correct to say that the overwhelming majority of equation-based simulations can be found in these domains, it does not do justice to the myriad of simulations also found in other scientific fields. To extend the examples, the Solow-Swan model of economic growth is a case in economy, and the Lotka-Volterra model of predator-prey works for sociology as well as for biology. 

As mentioned, this class of simulations implement mathematical models on the computer. But, is that simply so? From section \ref{sec:What-are-CS?} we learned that this way of thinking corresponds to the problem-solving viewpoint of computer simulation. According to its advocates, there is little that mediates between the mathematical model and its implementation on the computer as a computer simulation. The opposite view, the decription of pattern of behavior viewpoint, takes it that there is in fact a methodology that facilitates the implementation of -- a multiplicity of -- mathematical models into the computer. To have a better grasp of a typical equation-based simulation -- and to determine which viewpoint is closer to the actual practices --, let us briefly analyze an example of a recent simulation on the age of the Sahara desert.

Zhongshi Zhang et al. believed that the Sahara desert emerged during the Tortonian stage -- approximately 7-11 million years ago -- of the Late Miocene epoch after a period of aridity in the north African region \cite{Zhang2014} . To prove their hypothesis, Zhang's team decided to simulate the climate change in these regions on geological timescales and over the past 30 million years. The Sahara's age, according to the simulation is, in effect, of about 7 to 11 million years old. With this result in hand, Zhang's et al. were able to oppose most of the standard estimations of the age of the Sahara, which take it to be about 2--3 million years old at the onset of the Quaternary ice ages. How did they actually simulate such a complex empirical system?

First, the authors do not implement one grand model of climate change on the computer and calculate it until obtaining the results. This is the problem-solving viewpoint that understand computer simulations solely as means for computation. Computer simulations are much like a laboratory workbench, where scientists subtly combine pieces of theory, with bits of data, and a lot of expertise knowledge and instinct. The process is in fact complex, messy, and in many cases non-standardized. Zhang's team made use of a family of models, each performing different tasks and representing a different aspect of the target system. They made use of low- and high-resolution versions of the Norwegian Earth System Model (NorESM-L) for accounting for the series of geologic epochs, and the Community Atmosphere Model version 4 (CAM4) as the atmosphere component of NorESM-L.  In fact, the NorESM-L model is in itself a hierarchy of small models -- or mere components of a larger model -- representing the land, sea ice, the ocean, etc.

There is, in truth, no grand model that could tell us about the age of the Sahara. Instead, a patchwork of models -- some known and well-established, some speculative -- laws, principles, data, and bits of theory is what conform Zhang team's simulation. This should come as no surprise, as it is generally assumed that there is no general theory that either underpins or guides computer simulations. Moreover, simulations typically include non-linguistic information, such as design decisions, possible model bias, identified uncertainties,  and ``not included in this model'' disclaimers. Bentsen et al., when describing CAM4, furnish a good example of such a disclaimer: ``[a]erosol indirect effects on mixed-phase and ice clouds (e.g. \cite{Hoose2010} are not included in the current version of CAM4-Oslo'' \cite[689]{Bentsen2013}). 

Despite their lack of full theoretical underpinning, these simulations are still highly reliable as they represent a specific target system and are typically validated by standard verification and validation methods (see section \ref{Verification_and_validation}). In this respect, Zhang et al. are constantly reminding us that the model performs well in simulating the pre-industrial climate, that CAM4 simulate the patterns of modern African rainfall reasonably well, and other similar confirmatory stances of the models. Such reminders, of course, cannot stop objections against the results of the simulation. In particular, critics of Zhang's work point out the lack of evidence for validating their results. Stefan Kr\"{o}pelin is a chief detractor of using computer simulations for these kinds of target system. He admits that, although the model is interesting, it is mainly  ``numerical speculation based on almost non-existent geological evidence (...) Nothing you can find in the Sahara is older than 500,000 years, and in terms of Saharan climate even our knowledge of the past 10,000 is full of gaps'' \cite{Kroepelin2006}. The response by Zhang et al. is that the evidence for the early onset of the Saharan aridity is highly contentious. Mathieu Schuster also disagrees with Kr\"{o}pelin's interpretation of data. According to him, ``although it is true that too little is known about the ancient geology of the region [...] the 2006 Chad study [...] as well as the ones that reported increases in dust and pollen from sediment, contained `strong pieces of evidence to support our new findings''' \cite{Schuster2006}. In fact, the simulation by Zhang and his team come to support some claims already in the literature. Anil Gupta and his team claim an increased upwelling in the Indian Ocean about 7 to 8 million years ago \cite{Gupta2004}; and Gilles Ramstein and his team used modeling experiments to show Eurasian summer temperatures increase in response to the Tethys shrinkage, which would also enhance the monsoon circulation \cite{Ramstein1997}.

To state that computer simulations are unreliable, or that their results do not correlate to the way the world is requires more than just a claim that there is no `evidence' that supports the unreliability of the simulation.\footnote{This is especially true for the traditional sense of `evidence' (i.e., empirically based) that Kr\"{o}pelin refers to. Other forms of evidence also include results of well-established simulations, verification and validation, convergence of solutions, etc.} Other indicators of reliability play a central role as well. For instance, the ability of the simulation to explain and predict direct or related phenomena. According to Zhang et al., their simulation shows that the African summer monsoon was drastically weakened by the Tethys Sea shrinking during the Tortonian stage, allowing the alteration of the mean climate of the region. Such climatic change, the researchers speculate, ``probably caused the shifts in Asian and African flora and fauna observed during the same period, with possible links to the emergence of early hominis in north Africa'' \cite[401]{Zhang2014}. Interestingly enough, researchers could reach such a conclusion only by means of running computer simulations.

Allow me to finish this section with a short description of the general computational methods for solving the equation-based simulations. Depending on the problem and the availability of resources, one or more of the following methods apply: \emph{analytical methods}, \emph{numerical analysis} and \emph{stochastic techniques}. 

\begin{itemize}
	\item \emph{Exact solutions:} this is the simplest method of all. It consists of carrying out the operations specified in the simulation in a similar fashion as a mathematician would do using pen and paper. That is, if the simulation consist in adding $2 + 2$, then the result must be $4$ -- as opposed to an approximate solution. Computers have the same capacity to find the exact solutions to certain operations as any other computational mechanism, including our brain. The efficacy of this method depends, however, on whether the size of the `word' in a computer is large enough for carrying out the mathematical operation.\footnote{A `word' represents the minimum unit of data used by a particular computer architecture. It is a fixed sized group of bits that are handled as a unit by the processor.} If the operation exceeds its size, then round-off and truncation mechanisms intervene for the operation to be possible, though with a loss in accuracy. 
	
	\item \emph{Computer-implemented numerical methods}: this method refers to computer-implemented methods for calculating the simulation model by approximation. Although mathematical studies on numerical analysis predate the use of computers, they gain importance with the introduction of computers for scientific and engineering purposes. These methods are used for solving PDE and ODE equations, and include linear interpolation, the Runge-Kutta method, the Adams-Moulton method, Lagrange interpolation polynomial, Gaussian elimination, and Euler's method, among others. Let it be noted that each method is used for solving a specific kind of PDE and ODE, depending on how many derivatives involve the unknown function of $n$ variables.
	
	\item \emph{Stochastic techniques:} for higher order dimensions, both exact solutions and computer-implemented numerical methods become prohibitively expensive in terms of computational time and resources. Stochastic techniques rely on methods that use pseudo-random numbers; that is, numbers generated by a numerical engine.\footnote{The prefix `pseudo' reflects the fact that these methods are based on an algorithm that produces numbers on a recursive basis, eventually repeating the series of numbers produced. Pure randomness in computers can never be achieved.} The most famous stochastic method is the Monte Carlo method, which is particularly useful for simulating systems with many coupled degrees of freedom such as fluids, disordered materials, strongly coupled solids, and cellular structures, to mention a few.\\
\end{itemize}

\section{Concluding remarks}

This chapter had the sole purpose of addressing the question `what is a computer simulation?' This is of course an important question, since it sets the grounds for much of what is discussed about computer simulations later in this book. For this reason, the first part of the chapter deals with some historical remarks about the many attempts to define computer simulations, whether offered by engineers, scientists or philosophers. In this context, I distinguished two kinds of definitions. Those that emphasize the computing power of computer simulations -- called the \textit{problem-solving technique} viewpoint -- and those that take computer simulations to have, as a chief feature, the capacity to represent a given target system -- called \textit{description of patterns of behavior} viewpoint. Although there are a handful of definitions where both viewpoints are combined, and arguably one that does not fit with our distinction, in general researchers across fields agree on conceptualizing computer simulations as one or the other viewpoint. 

The second part of this chapter dealt with three different kinds of computer simulations, as standardly found in the literature. These are, \textit{cellular automata}, \textit{agent-based simulations}, and \textit{equation-based simulations}. As warned, this is neither an exhaustive taxonomy nor offers a unique classification. It could be relatively simple to show how an agent-based simulation could be interpreted as cellular automata (e.g., when focused on their nature as agents/cells), or as an equation-based simulation (e.g., if the inner structure of an agent are equations). The key is to see which characteristic of the computer simulation is highlighted. Here, I offer some criteria for a sound characterization of each type. A final warning is issued, however, regarding the methodology and epistemology tailored to each kind. It is not difficult to show that each kind of computer simulation entails specific and distinct methodological and epistemological concerns, and therefore they require a different treatment in their own way. In the reminder of this book, I focus my attention solely on the so-called equation-based simulations.

\bibliography{./References_Springer}

\begin{thebibliography}{10}

\bibitem{Ajelli2010}
Marco Ajelli, Bruno Gon{\c c}alves, Duygu Balcan, Vittoria Colizza, Hao Hu,
  Jos{\'e}~J. Ramasco, Stefano Merler, and Alessandro Vespignani.
\newblock Comparing {{Large}}-{{Scale Computational Approaches}} to {{Epidemic
  Modeling}}: {{Agent}}-{{Based}} versus {{Structured Metapopulation Models}}.
\newblock 10(190):1--13.

\bibitem{Ardourel2017}
Vincent Ardourel and Julie Jebeile.
\newblock On the presumed superiority of analytical solutions over numerical
  methods.
\newblock 7(2):201--220.

\bibitem{Banks2010}
Jerry Banks, John Carson, Barry~L. Nelson, and David Nicol.
\newblock {\em Discrete-Event System Simulation}.
\newblock Prentice Hall.

\bibitem{Barnstorff2010}
Kathy Barnstorff.
\newblock {\em X-{{51A Makes Longest Scramjet Flight}}}.
\newblock 2010.

\bibitem{Beisbart2012}
Claus Beisbart.
\newblock How can computer simulations produce new knowledge?
\newblock {\em European Journal for Philosophy of Science}, 2:395--434, 2012.

\bibitem{Bentsen2013}
M.~Bentsen, I.~Bethke, J.~B. Debernard, T.~Iversen, A.~Kirkev{\aa}g, ?.~Seland,
  H.~Drange, C.~Roelandt, I.~A. Seierstad, C.~Hoose, and J.~E.
  Kristj{\'a}nsson.
\newblock The {{Norwegian Earth System Model}}, {{NorESM1}}-{{M}} \textendash{}
  {{Part}} 1: {{Description}} and {{Basic Evaluation}} of the {{Physical
  Climate}}.
\newblock 6(3):687--720.

\bibitem{Birtwistle1979}
G.~M. Birtwistle.
\newblock {\em {DEMOS} A System for Discrete Event Modelling on Simula}.
\newblock The MacMillan Press, 1979.
\newblock (Reprint 2003).

\bibitem{Ceruzzi1998}
Paul~E. Ceruzzi.
\newblock {\em A History of Modern Computing}.
\newblock {MIT Press}, 1998.

\bibitem{Colburn2007}
Timothy Colburn and Gary Shute.
\newblock Abstraction in computer science.
\newblock {\em Minds and Machines}, 17(2):169--184, June 2007.

\bibitem{DeMol2014}
Liesbeth De~Mol and Giuseppe Primiero.
\newblock Facing computing as technique: towards a history and philosophy of
  computing.
\newblock {\em Philosophy \& Technology}, 27(3):321--326, 2014.

\bibitem{DeMol2015}
Liesbeth De~Mol and Giuseppe Primiero.
\newblock When logic meets engineering: introduction to logical issues in the
  history and philosophy of computer science.
\newblock {\em History and Philosophy of Logic}, 36(3):195--204, 2015.

\bibitem{Drogoul1992}
Alexis Drogoul, Jacques Ferber, and Christophe Cambier.
\newblock Multi-agent simulation as a tool for modeling societies: Application
  to social differentiation in ant colonies.
\newblock In G.~Nigel Gilbert, editor, {\em Simulating {{Societies}}}, pages
  49--62. Guilford : University of Surrey, 1992.

\bibitem{Duranunderreview}
Juan~M. Dur{\'a}n.
\newblock The novelty of computer simulations: new challenges for the
  philosophy of science.
\newblock {\em under review}, under review.

\bibitem{ElSkaf2013}
Rawad El~Skaf and Cyrille Imbert.
\newblock Unfolding in the empirical sciences: Experiments, thought experiments
  and computer simulations.
\newblock {\em Synthese}, 190(16):3451--3474, 2013.

\bibitem{Keller2003}
Evelyn Fox~Keller.
\newblock Models, simulations, and "computer experiments".
\newblock In Hans Radder, editor, {\em The Philosophy of Scientific
  Experimentation}, pages 198--215. University of Pittsburgh Press, 2003.

\bibitem{Frigg2009a}
Roman Frigg and Julian Reiss.
\newblock The philosophy of simulation: Hot new issues or same old stew?
\newblock 169(3):593--613.

\bibitem{Gardner1970}
Martin Gardner.
\newblock The fantastic combinations of john conway's new solitaire game
  ``life''.
\newblock {\em Scientific American}, 223(4):120--123, 1970.

\bibitem{Gilbert2005}
G.~Nigel Gilbert and Klaus~G. Troitzsch.
\newblock {\em Simulation for the {{Social Scientist}}}.
\newblock {Open University Press}, Maidenhead, England ; New York, NY, 2nd ed
  edition, 2005.
\newblock LCCB: HM51 .G54 2005.

\bibitem{Guala2002}
Francesco Guala.
\newblock Models, simulations, and experiments.
\newblock In L.~Magnani and N.~J. Nersessian, editors, {\em Model-{{Based
  Reasoning}}: {{Science}}, {{Technology}}, {{Values}}}, pages 59--74. Kluwer
  Academic, 2002.

\bibitem{Gupta2004}
Anil~K. Gupta, Raj~K. Singh, Sudheer Joseph, and Ellen Thomas.
\newblock Indian {{Ocean High}}-{{Productivity Event}} (10\textendash{}8
  {{Ma}}): {{Linked}} to {{Global Cooling}} or to the {{Initiation}} of the
  {{Indian Monsoons}}?
\newblock {\em Geology}, 32(9):753, 2004.

\bibitem{Hartmann1996}
Stephan Hartmann.
\newblock The world as a process: {{Simulations}} in the natural and social
  sciences.
\newblock In R.~Hegselmann, Ulrich Mueller, and Klaus~G. Troitzsch, editors,
  {\em Modelling and {{Simulation}} in the {{Social Sciences}} from the
  {{Philosophy}} of {{Science Point}} of {{View}}}, pages 77--100. Springer.

\bibitem{Himeno2013}
Ryutaro Himeno.
\newblock Largest {{Neuronal Network Simulation Achieved Using K Computer}}.

\bibitem{Hoose2010}
Corinna Hoose, J{\'o}n~Egill Kristj{\'a}nsson, Jen-Ping Chen, and Anupam Hazra.
\newblock A {{Classical}}-{{Theory}}-{{Based Parameterization}} of
  {{Heterogeneous Ice Nucleation}} by {{Mineral Dust}}, {{Soot}}, and
  {{Biological Particles}} in a {{Global Climate Model}}.
\newblock {\em Journal of the Atmospheric Sciences}, 67(8):2483--2503, August
  2010.

\bibitem{Humphreys1990}
Paul~W. Humphreys.
\newblock Computer simulations.
\newblock 2:497--506.

\bibitem{Humphreys2004}
Paul~W. Humphreys.
\newblock {\em Extending {{Ourselves}}: {{Computational Science}},
  {{Empiricism}}, and {{Scientific Method}}}.
\newblock Oxford University Press.

\bibitem{Humphreys2009}
Paul~W. Humphreys.
\newblock The philosophical novelty of computer simulation methods.
\newblock 169(3):615--626.

\bibitem{Kroepelin2006}
S.~Kroepelin.
\newblock Revisiting the {{Age}} of the {{Sahara Desert}}.
\newblock 312(5777):1138b--1139b, May 2006.

\bibitem{Laboratory2015}
Los Alamos~National Laboratory.
\newblock {\em Largest {{Computational Biology Simulation Mimics Life}}'s
  {{Most Essential Nanomachine}}}.

\bibitem{Lesne2007}
Annick Lesne.
\newblock The discrete versus continuous controversy in physics.
\newblock {\em Mathematical Structures in Computer Science}, 17(2):185--223,
  2007.

\bibitem{McMillan1965}
Claude McMillan and Richard~F Gonz{\'a}lez.
\newblock {\em Systems {{Analysis}}: {{A Computer Approach}} to {{Decision
  Models}}}.
\newblock Irwin.

\bibitem{Morrison2009}
Margaret Morrison.
\newblock Models, measurement and computer simulation: {{The}} changing face of
  experimentation.
\newblock 143(1):33--57.

\bibitem{Morrison2015}
Margaret Morrison.
\newblock {\em Reconstructing {{Reality}}. {{Models}}, {{Mathematics}}, and
  {{Simulations}}}.
\newblock {Oxford University Press}, 2015.

\bibitem{Naylor1967}
Thomas~H.. Naylor, J.~M. Finger, James~L. McKenney, Williams~E. Schrank, and
  Charles~C. Holt.
\newblock Verification of {{Computer Simulation Models}}.
\newblock {\em Management Science}, 14(2):92--106, 1967.

\bibitem{Oberkampf2003}
William~L Oberkampf, Timothy~G Trucano, and Charles Hirsch.
\newblock Verification, validation, and predictive capability in computational
  engineering and physics.

\bibitem{Oeren1984}
Tuncer \"Oren.
\newblock Model-based activities: A paradigm shift.
\newblock In Tuncer \"Oren, B.~P Zeigler, and M.~S. Elzas, editors, {\em
  Simulation and Model-Based Methodologies: An Integrative View}, pages 3--40.
  Springer, 1984.

\bibitem{Oeren2011a}
Tuncer \"Oren.
\newblock A critical review of definitions and about 400 types of modeling and
  simulation.
\newblock {\em SCS M\&S Magazine}, 2(3):142--151, 2011.

\bibitem{Oeren2011}
Tuncer \"Oren.
\newblock The many facets of simulation through a collection of about 100
  definitions.
\newblock {\em SCS M\&S Magazine}, 2(2):82--92, 2011.

\bibitem{Oeren1982}
Tuncer \"Oren, B.~P. Zeigler, and M.~S. Elzas, editors.
\newblock {\em Proceedings of the NATO Advanced Study Institute on Simulation
  and Mode-Based Methodologies: An Integrative View}. NATO ASI Series.
  Springer-Verlag, 1982.

\bibitem{Parker2009}
Wendy~S Parker.
\newblock Does matter really matters? {{Computer}} simulations, experiments,
  and materiality.
\newblock {\em Synthese}, 169(3):483--496, 2009.

\bibitem{Ramstein1997}
Gilles Ramstein, Fr{\'e}d{\'e}ric Fluteau, Jean Besse, and Sylvie Joussaume.
\newblock Effect of {{Orogeny}}, {{Plate Motion}} and {{Land}}\textendash{}sea
  {{Distribution}} on {{Eurasian Climate Change}} over the {{Past}} 30
  {{Million Years}}.
\newblock 386(6627):788--795, April 1997.

\bibitem{Saam2016}
Nicole~J. Saam.
\newblock What is a {{Computer Simulation}}? {{A Review}} of a {{Passionate
  Debate}}.
\newblock {\em Journal for General Philosophy of Science}, pages 1--17, 2016.

\bibitem{Schelling1971}
Thomas~C. Schelling.
\newblock On the {{Ecology}} of {{Micromotives}}.
\newblock {\em National Affairs}, 25(Fall), 1971.

\bibitem{Schuster2006}
M.~Schuster.
\newblock The age of the sahara desert.
\newblock {\em Science}, 311(5762):821--821, February 2006.

\bibitem{Shannon1975}
Robert~E. Shannon.
\newblock {\em Systems Simulation: The Art and Science}.
\newblock Prentice Hall, Englewood Cliffs, N.J, 1975.

\bibitem{Shannon1978}
Robert~E. Shannon.
\newblock Design and analysis of simulation experiments.
\newblock In {\em Proceedings of the 10th Conference on Winter Simulation -
  Volume I}, pages 53--61. IEEE Press, 1978.

\bibitem{Shannon1998}
Robert~E. Shannon.
\newblock Introduction to the art and science of simulation.
\newblock In {\em Proceedings of the 30th Conference on Winter Simulation},
  pages 7--14, Los Alamitos, CA, USA, 1998. {IEEE Computer Society Press}.

\bibitem{Shubik1960}
Martin Shubik.
\newblock Simulation of the {{Industry}} and {{The Firm}}.
\newblock {\em The American Economic Review}, 50(5):908--919, 1960.

\bibitem{Smith1981}
Reid~G. Smith and Randall Davis.
\newblock Frameworks for cooperation in distributed problem solving.
\newblock {\em IEEE Transactions on Systems, Man, and Cybernetics},
  11(1):61--70, 1981.

\bibitem{Teichroew1966}
Daniel Teichroew and John~Francis Lubin.
\newblock Computer simulation -{{Discussion}} of the technique and comparison
  of languages.
\newblock 9(10):723--741.

\bibitem{Toffoli1984}
Tommaso Toffoli.
\newblock {{CAM}}: {{A}} high-performance cellular-automaton machine.
\newblock {\em Physica D: Nonlinear Phenomena}, 10(1-2):195--204, 1984.

\bibitem{Vallverdu2014}
Jordi Vallverd{\'u}.
\newblock What are simulations? an epistemological approach.
\newblock {\em Procedia Technology}, 13:6--15, 2014.

\bibitem{Vichniac1984}
G??rard~Y. Vichniac.
\newblock Simulating physics with cellular automata.
\newblock {\em Physica D: Nonlinear Phenomena}, 10(1-2):96--116, 1984.

\bibitem{Weisberg2013}
Michael Weisberg.
\newblock {\em Simulation and {{Similarity}}}.
\newblock {Oxford University Press}.

\bibitem{Winsberg2010}
Eric Winsberg.
\newblock {\em Science in the {{Age}} of {{Computer Simulation}}}.
\newblock {University of Chicago Press}.

\bibitem{Winsberg2015}
Eric Winsberg.
\newblock Computer simulations in science.
\newblock In Edward~N. Zalta, editor, {\em The Stanford Encyclopedia of
  Philosophy}. 2015.

\bibitem{Wolfram1984a}
Stephen Wolfram.
\newblock Preface.
\newblock {\em Physica 10D}, pages vii--xii, 1984.

\bibitem{Wolfram1984}
Stephen Wolfram.
\newblock Universality and {{Complexity}} in {{Cellular Automata}}.
\newblock {\em Physica D: Nonlinear Phenomena}, pages 1--35, 1984.

\bibitem{Woolfson1999a}
Michael~M. Woolfson and Geoffrey~J. Pert.
\newblock {\em An {{Introduction}} to {{Computer Simulations}}}.
\newblock {Oxford University Press}.

\bibitem{Woolfson1999}
Michael~M. Woolfson and Geoffrey~J. Pert.
\newblock {\em {{SATELLIT}}.{{FOR}}}.

\bibitem{Zhang2014}
Zhongshi Zhang, Gilles Ramstein, Mathieu Schuster, Camille Li, Camille Contoux,
  and Qing Yan.
\newblock Aridification of the sahara desert caused by tethys sea shrinkage
  during the late miocene.
\newblock {\em Nature}, 513(7518):401--404, September 2014.

\end{thebibliography}
\bibliographystyle{plain}

\end{document}